%% file: main.tex
\documentclass[amsmath,onecolumn, floatfix,nofootinbib, prx, aps, usenames, dvipsnames,11pt]{quantumarticle}
\pdfoutput=1

\input{macros}

\begin{document}

\title{Thought experiments in a quantum computer}

\author{Nuriya Nurgalieva}
\affiliation{Institute for Theoretical Physics, ETH Z\"{u}rich, 8093 Z\"{u}rich, Switzerland}

\author{Simon Mathis}
\affiliation{Institute for Theoretical Physics, ETH Z\"{u}rich, 8093 Z\"{u}rich, Switzerland}

\author{Lídia del Rio}
\affiliation{Institute for Theoretical Physics, ETH Z\"{u}rich, 8093 Z\"{u}rich, Switzerland}
\affiliation{Quantum Center, ETH Zurich, 8093 Z\"{u}rich, Switzerland}

\author{Renato Renner}
\affiliation{Institute for Theoretical Physics, ETH Z\"{u}rich, 8093 Z\"{u}rich, Switzerland}

\date{}

\begin{abstract}

We introduce a software package that allows users to design and run simulations of thought experiments in quantum theory. In particular, it covers cases where several reasoning agents are modelled as quantum systems, such as Wigner's friend experiment. Users can customize the protocol of the experiment, the inner workings of agents (including a quantum circuit that models their reasoning process), the abstract logical system used (which may or not allow agents to combine premises and make inferences about each other's reasoning), and the interpretation of quantum theory used by different agents.
Our open-source software is written in a quantum programming language, ProjectQ, and runs on classical or quantum hardware. As an example, we model the Frauchiger-Renner extended Wigner's friend thought experiment, where agents are allowed to measure each other's physical memories, and make inferences about each other's reasoning.
\end{abstract}

\maketitle

\vspace{-1cm}

\setlength{\epigraphwidth}{5in}
\epigraph{Truth is a matter of the imagination.}{Ursula K. Le Guin, \emph{The Left Hand of Darkness}}

\begin{quote}
    Software available at  \url{https://github.com/jangnur/Quanundrum}  \cite{Nurgalieva_Quanundrum_2021}. 
\end{quote}


\begin{figure}[H]
\centering
\includegraphics[scale=0.17]{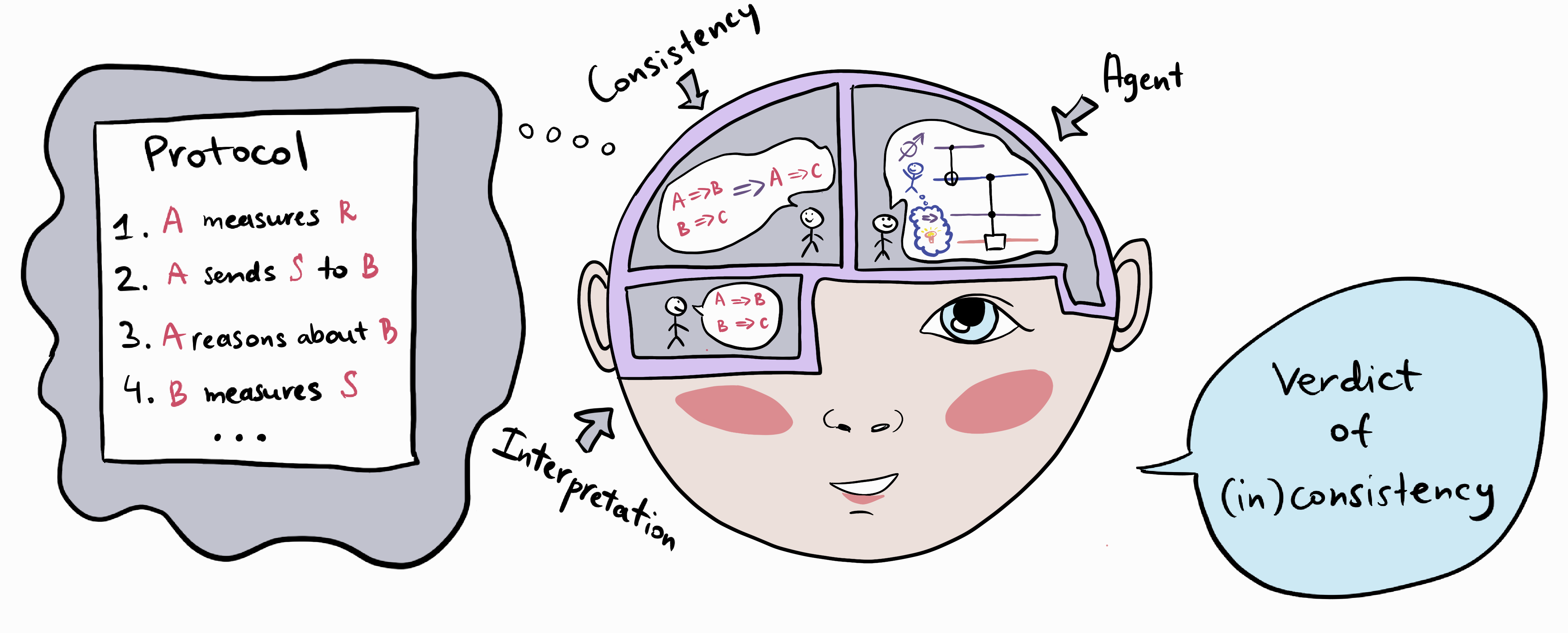}
\caption{{ \bf \textit{Quanundrum} software structure.} Our package can model quantum multi-agent scenarios where agents reason about each other's experiments and thoughts. Its modular components customize the experimental setting \emph{(Protocol module)}, the inferences allowed given experimental observations \emph{(Interpretation module)}, how those inferences are logically combined and propagated \emph{(Consistency)}, and how agents' brains are physically modeled \emph{(Agent)}. The program outputs the logical conclusions of all agents and  whether they are compatible.}
\label{fig:structure}
\end{figure}

\section{Introduction}
\label{sec:introduction}
\input{introduction}

\section{Physical models of agents: memory and reasoning}
\label{sec:memory}
\input{memory}

\section{Abstract reasoning: interpretation and logic}
\label{sec:logic-inter}
\input{logic-inter}

\section{Discussion}
\label{sec:conclusions}
\input{conclusions}

\subsection*{Acknowledgements} We thank M. Iazzi and several anonymous conference referees for useful feedback.
We acknowledge support from the Swiss
National Science Foundation through 
SNSF project No.\ $200021\_188541$ and through the
the National Centre of
Competence in Research \emph{Quantum Science and Technology}
(QSIT). 
LdR further acknowledges support from  the FQXi grant \emph{Consciousness in the Physical World}.  LdR is grateful for the hospitality of Perimeter Institute where part of this work was carried out. Research at Perimeter Institute is supported in part by the Government of Canada through the Department of Innovation, Science and Economic Development and by the Province of Ontario through the Ministry of Colleges and Universities.

\subsection*{Author contributions}
This project started as an extra-curricular project of SM, back then a master student, proposed and supervised by the remaining authors. All authors contributed equally for the ideas in the paper and software. The software was designed by SM with input from the remaining authors, and later revised by NN. All Jupyter notebooks and figures were produced by NN. NN and LdR wrote  this manuscript. LdR would like a cookie for drawing the circuits, but will settle for a more compliant alternative to QCircuit.

\newpage
\appendix


\section*{\sc{Appendix}}
\section{Examples of simple protocols}

\label{appendix:examples}
\input{appendix_examples}

\section{Application: the Frauchiger-Renner thought experiment}
\label{sec:FR}

\input{FR_protocol}

\vspace{4cm}
\bibliographystyle{unsrtnat}
\input{main.bbl} 

\end{document}

%% file: macros.tex
\usepackage[utf8]{inputenc}
\usepackage[english]{babel}
\usepackage[T1]{fontenc}

\usepackage{mathtools}
\usepackage{enumitem}

\usepackage[usenames,dvipsnames]{xcolor}
\usepackage{graphicx}
\usepackage[ colorlinks = true, 
             linkcolor = MidnightBlue,
             urlcolor  = MidnightBlue,
             citecolor = orange,
             anchorcolor = ForestGreen,
]{hyperref} 
\usepackage{float}

\usepackage[sans]{dsfont} 
\usepackage{bbm}
\usepackage[mathscr]{euscript}
\usepackage{cancel} 
\usepackage{amsfonts}

\usepackage{tcolorbox}
\tcbuselibrary{theorems} 

\usepackage{amsmath}
\usepackage{amssymb}
\usepackage{units}
\usepackage{framed}
\usepackage{mdframed}
\usepackage{thmtools, thm-restate} 
\usepackage{ucs}
\usepackage{subcaption}
\usepackage{fullpage}
\usepackage{csquotes}
\usepackage{epigraph}

\usepackage{multicol}
\usepackage{enumerate}

\usepackage{tikz}
\usetikzlibrary{arrows}
\usepackage{rotating, xcolor}
\usetikzlibrary{shapes}
\usetikzlibrary{hobby}
\usetikzlibrary{decorations.markings}

\usepackage{caption}
\makeatletter
\def\l@subsubsection#1#2{}
\def\l@subsection#1#2{}
\makeatother

\tcbset{colback=orange!5,colframe=orange!75!yellow, 
fonttitle= \bfseries, float=htb}

	\newcommand{\red}[1]{\textcolor{Red}{#1}}
	\newcommand{\blue}[1]{\textcolor{Blue}{#1}}
	\newcommand{\green}[1]{\textcolor{OliveGreen}{#1}}

    \newcommand{\ket}[1]{\vert  #1 \rangle}

	\renewcommand{\vec}[1]{\mathbf{#1}}

\newcommand{\footnoterecall}[1]{\hyperref[#1]{\footnotemark[\value{#1}]}}

\usepackage[page]{totalcount}
\usepackage{listings}
\usepackage{qcircuit}

\definecolor{dblue}{rgb}{0,0,0.7}
\definecolor{lgray}{rgb}{.4,.4,.4}
\usetikzlibrary{arrows,decorations.markings}

%% file: introduction.tex
Let us summarize the main contributions of this work immediately, before describing the motivation in detail in Section \ref{section:intro:motivation}.

\subsection{Contribution of this work}
\label{section:intro:contributions}

\paragraph{A modular tool to test multi-agent quantum thought experiments.} We introduce a software package, \emph{Quanundrum}, to run  quantum mechanical thought experiments where agents are modeled as quantum systems and have to reason about each other's  experimental results (Figure~\ref{fig:structure}).
Users can customize the  following modules: 
\begin{itemize}
    \item {\bf Protocol module:} specifies the experimental setting, including number of agents and other quantum systems, measurements and physical transformations carried out by different agents, and which chains of inferences we are interested in analysing. 
    \item {\bf Agent module:}  specifies agents' physical memories and processors, defining how abstract reasoning is implemented as a physical process, e.g.\ as a quantum circuit. 
    \item {\bf Interpretation module:} defines  the immediate inferences drawn from an experimental outcome, which depends on the interpretation of quantum theory applied by an agent. 
    \item {\bf Consistency module:} defines the axioms of abstract logic that determine how inferences are combined and how knowledge is propagated into complex reasoning; it also determines the subjects that agents are allowed to reason about. 
\end{itemize}

\paragraph{Structure of this manuscript.} This paper introduces the conceptual idea of the software and its main components. In Section~\ref{section:intro:motivation} we motivate the need for  software to simulate quantum thought experiments.  In Section~\ref{sec:memory}, we detail how we physically model agents' memories and reasoning processes. In Section~\ref{sec:logic-inter}, we look at how agents's inferences are propagated and combined to complex reasoning. We provide an application to the Frauchiger-Renner thought experiment in Section~\ref{sec:FR}, and examples of simpler protocols in Appendix~\ref{appendix:examples}. We discuss the implications of this work in Section~\ref{sec:conclusions}. For documented step-by-step examples and customization, we recommend assessing the software directly.

\paragraph{Structure of the software repository.} 
The open-source software package \textit{Quanundrum} is publicly available~\cite{Nurgalieva_Quanundrum_2021}. The software is written in ProjectQ~\cite{Steiger2018}, a free open-source quantum programming language based on Python.
In the repository, one can find: installation instructions; instructions on how to run and customize the software; Jupyter notebooks with examples: simple test scenarios and the Frauchiger-Renner thought experiment~\cite{Frauchiger2018} for two different interpretations; explanations of the experimental settings and conclusions; descriptions of the physical modelling of rational agents as small quantum computers; and descriptions of the implementation of  interpretations.

\subsection{Motivation}
\label{section:intro:motivation}

\paragraph{Thought experiments in physics.} In science and philosophy, thought experiments  use an imaginary setting to draw conclusions about our physical reality  --- in  regimes that are technologically or ethically unfeasible. For example  Maxwell's demon~\cite{Maxwell1871} thought experiment, which explores what would happen if a being could access microscopic degrees of freedom of a gas --- something unreachable at the time ---  led to  conceptual and technological breakthroughs. The conclusion of the thought experiment was a seeming violation of the second law of thermodynamics; inspired by this inconsistency, Landauer's principle~\cite{Landauer1961} was developed, stating a minimum possible amount of energy required to erase one bit of information to save the second law; the connections between information theory and thermodynamics, with multiple applications in the quantum regime, stem from this insight. Other famous examples include (but are not limited to) the twins paradox in special relativity~\cite{Einstein1911}, Fermi's  paradox in cosmology, and, in  quantum theory, Schrödinger's cat~\cite{Schroedinger1926} and Wigner's friend~\cite{Wigner1961} scenarios.

\paragraph{Analysing thought experiments.}  Thought experiments have a similar structure to the planning of actual physical experiments: a particular situation in the scope of a theory is visualized; an operation or a set of operations is carried out; we see what happens; finally, we draw a conclusion. Thought experiments can be analised and criticized  from different angles: perhaps the setting is not achievable in reality, and the experiment cannot be scaled to more realistic settings, rendering the conclusions inconsequential; perhaps there are hidden assumptions that weaken the conclusions of the experiment; sometimes the different assumptions are coarse-grained in ways that make it hard to identify the critical one. To achieve a deep understanding of the consequences of a thought experiment, it is crucial to be able to specify in a modular way the different components that go into it: what theory is used, what is the experimental setting, what are the assumptions on the underlying logical system, what relevant parts of the experiment could we be ignoring? For example, in the case of Maxwell's demon, it was crucial to realize that the demon's knowledge about the position and momentum of gas particles had to be physically stored \emph{somewhere} (the demon's memory), which had been overlooked in the original proposal of the experiment.

\subsection{Quantum thought experiments as programs}

\paragraph{Philosophical motivation: testing quantum theory and interpretations.} In quantum mechanics, thought experiments are used to prove theorems that tell us more about the underlying structure of quantum theory, for example by ruling out different interpretations and properties of the theory \cite{Hardy1993, Brukner2018, Wigner1961, Kochen1967}. Modelling these thought experiments as computer programs forces us to rigorously specify what a given interpretation entails, and to test its consequences under different scenarios. It shifts the focus from sometimes vague philosophical considerations to concrete and mathematically complete descriptions, which can be modelled explicitly. It can also allow us to understand which interpretations result in distinct physical theories, and which affect only the abstract reasoning of agents. Finally, once an interpretation is implemented in code, it can be more easily tweaked and re-tested in critical settings, which leads to faster development of consistent interpretations.   

\paragraph{Practical motivation: quantum computing networks.} From a more practical perspective, we can envision a future where we would like to set up networks of quantum computers, and program them to reason independently about events happening elsewhere in the network. The reasoning these node computers conduct should be logically \textit{consistent}, akin to the classical reasoning we are used to. For example, the computers should not reach conflicting conclusions with regards to measurements' outcomes, or should be free to conclude $A \Rightarrow C$ from knowing the implications $A \Rightarrow B$ and $B \Rightarrow C$.\footnote{$A, B$ and $C$ here are any statements that can be made in such a quantum computer network; for example, ``the outcome of the measurement of a qubit $S$ is $a=1$'', or ``it is going to rain tomorrow''. All the inferences are assumed to be deterministic, that is, they hold true with probability 1.}

\paragraph{A yellow flag for logical quantum networks.} 
Surprisingly, these simple logical requirements can prove to be difficult to achieve in a quantum network. Taking one specific quantum mechanical example, in a thought experiment introduced by Frauchiger and Renner in~\cite{Frauchiger2018} (in the following \textit{FR thought experiment}) agents, who are allowed to reason about each other's outcomes and are also modelled as quantum systems, come to a logical contradiction.\footnote{We detail this scenario  in Section~\ref{sec:FR}.}  More generally, we have shown that the traditional framework for agents reasoning in a multi-agent setting, namely, classical modal logic~\cite{Kripke2007}, fails in quantum settings~\cite{NL2018}. These conclusions are  based on  assumptions on how agents are modelled, how they make predictions and inferences, and how these predictions and inferences are combined. 


\paragraph{Testing the boundaries of quantum network reasoning.}
This illustrates how thought experiments are crucial to determine the boundaries of the reasoning programmed into nodes. For example, a popular conservative view~\cite{Aaronson2018} is that, in order to avoid make false predictions, a local quantum computer must not be allowed to reason about global protocols that include a future measurement of its own memory. While this is not problematic for traditional quantum algorithms, which typically end with a measurement, it could severely restrict the power and range of predictions allowed to fully quantum computer networks. We hope that by carefully testing different scenarios we can find  weaker restrictions that avoid contradictions.

%% file: memory.tex
\paragraph{Classical multi-agent logic: a light example.} Three logicians walk into a bar. ``Does everyone want a beer?'' the bartender asks. Alice answers ``I don't know.'' Then  Bob  says ``I don't know either.'' Finally, Charlie shouts ``Yes!'' -- and everyone gets their beer~\cite{LogiciansBeer}. Charlie's reasoning is simple: they consider Alice's point of view, and deduce that if Alice did not want a beer, she would have answered with a definitive ``No'', as not \textit{everyone} would have wanted a beer. Considering then Bob's viewpoint, Charlie concludes that Bob also deduced that Alice wants a beer, and that his reason for answering ``I don't know''  was that Bob did want a beer, but was unsure of Charlie's wishes. This allows Charlie to correctly assess that both of their colleagues want a beer, and, given that they also very much want a beer, correctly answer ``Yes!''

\paragraph{Physical modelling of memories, observation and reasoning.} In the previous example, Charlie observes their setting for a period of time, and reasons about their colleagues' reasoning based on those observations, and their internal model of how their colleagues reason (for example, taking for granted that they are very rational and literal logicians).
To model the physical implementation of this situation, we'd have to specify explicitly: what physical systems implement their abstract memories, how they write down observations into their memory, and how they reason about these statements given the theory they are allowed to apply.
If we were modelling agents in the context of quantum theory, we could start by implementing their memories as quantum systems represented by Hilbert spaces, and their reasoning processes as quantum channels on those Hilbert spaces and those of the systems observed. In particular we can consider a dilation of all the relevant systems so that we can look at unitary implementations of observation and reasoning processes. This will be convenient to later model them as quantum circuits. We will detail in the following how we model these different aspects in the quantum case, for very simple idealizations.

\subsection{Physical implementation of measurements and memory update rules}

\begin{figure}[t]
\centering
\begin{subfigure}{0.45\textwidth}
    \includegraphics[scale=0.28]{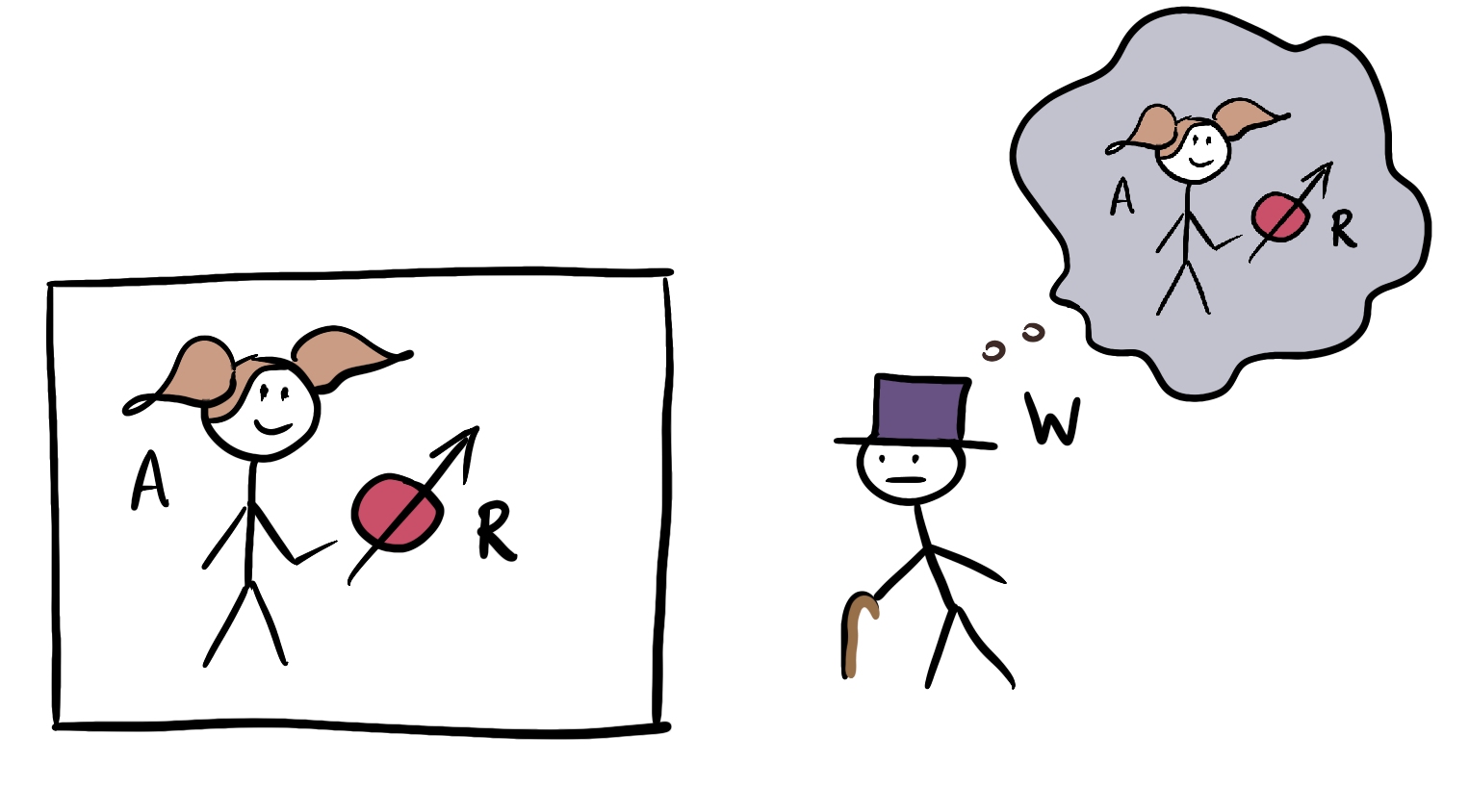}
    \caption{{Thought experiment setup.} }
        \label{fig:wigner-alice}
\end{subfigure}
\qquad \qquad
\begin{subfigure}{0.3\textwidth}
        \Qcircuit @C=1.3em @R=1.3em{ 
        \push{\hspace{6em}} &\lstick{\frac{1}{\sqrt 2}\ket 0_R + \frac{1}{\sqrt 2}\ket 1_R} & \meter & \rstick{\blue{a}} \cw \\
     &  \\ & \\ 
        &\lstick{ \frac{1}{\sqrt 2}\ket 0_R + \frac{1}{\sqrt 2}\ket 1_R} & \ctrl{1} & \qw \\
         &\lstick{\ket{0}_A} & \targ &  \qw \\
        }

    \caption{{Perspectives of Alice (top) and Wigner (bottom).}}
        \label{fig:wigner-circuit}
\end{subfigure}
\caption{{\bf Wigner's friend thought experiment~\cite{Wigner1961}.} An agent Alice holds in her lab a qubit $R$, initially in state $\frac{1}{\sqrt{2}}(\ket{0}_R+\ket{1}_R)$. Alice measures $R$ in  basis $\{\ket{0}_R,\ket{1}_R\}$, and records the outcome ($0$ or $1$) into her memory $A$, initially in the state $\ket 0_A$. Outside her lab stands a second agent, Wigner, who reasons about the state of the lab.  Alice subjectively observes a single outcome $a=0$ or $a=1$ and models the measurement as an irreversible transformation. In the neo-Copenhagen interpretation, Wigner can model Alice's lab as an isolated quantum system, and the joint evolution of $R$ and her memory $A$ as a unitary transformation: a  CNOT gate that coherently copies the measurement outcome  to Alice's memory $A$. 
}
\label{fig:wigner}
\end{figure}

\paragraph{Quantum measurements in the lab.} We provide a broad characterization of memory updates in arbitrary physical theories in~\cite{Vilasini2019}; here we illustrate our approach through simple examples.
To measure the spin of the particle in the lab, one can use for example the Stern-Gerlach setup. To experimentally realize it, we tune a magnetic field that  couples the  spin to one of the position degrees of freedom of the particle. The interaction Hamiltonian can be simplified to
\begin{gather*}
    \hat H_{int} = g\  \hat S_{Z_{\text{spin}}} \otimes \hat X_{\text{position}} , \qquad
    \mathcal H_{\text{particle}} =  \mathcal H_{\text{spin}} \otimes  \mathcal H_{\text{position}}, \qquad g\in \mathbbm R,
\end{gather*}
where $\hat S_Z$ is the spin observable and   $\hat X$ is the position operator. 
 This results in a momentum kick to the particle, whose  trajectory becomes entangled  with its spin; this can be macroscopically observed by installing a screen on the particle's path. In other settings, an external degree of freedom, like a pointer, takes the place of the position, so that measuring a  quantum system corresponds to entangling it with the pointer.

 \paragraph{Memories as entangled pointers.}
 Similarly, we can model an agent's memory as a degree of freedom akin to the pointer, which after the measurement becomes coherently correlated with the measured system. In a simplified toy example, both the relevant part of the memory $M$ and the measured system $S$ are qubits: the memory states $\ket 0_M$ and $\ket 1_M$ can encode the observed outcomes $0$ and $1$ respectively. Then, for a  measurement of $S$ in the computational basis $\{\ket 0_S, \ket 1_S\}$, the memory update describing the process of the agent ``observing the outcome and writing it down to their memory'' could be modelled as a CNOT gate between $S$ and $M$.  \footnote{There is a discussion to be held about the role of decoherence in having the subjective experience of seeing a single outcome. This is outside the scope of this paper, but we encourage it as a research direction.}

\paragraph{Subjective view of measurements.}  Wigner's friend \cite{Wigner1961} thought experiment (Figure~\ref{fig:wigner}) is a canonical example for how agents can subjectively model each other's measurements and memory update processes. 
An agent Alice holds in her lab a qubit $R$, initially in state $\frac{1}{\sqrt{2}}(\ket{0}_R+\ket{1}_R)$. Alice measures $R$ in the basis $\{\ket{0}_R,\ket{1}_R\}$, and records the outcome ($0$ or $1$) in her memory $A$. From Alice's perspective, the joint state of $R$ and her memory $A$ after the measurement is either $\ket0_R\ket0_A$ or $\ket1_R\ket1_A$. Outside her lab stands a second agent, Wigner, who models Alice's measurement as a von Neumann interaction scheme~\cite{vonNeumann1955}, through which the state or the observed system $R$ is coherently copied to the memory $A$. 
In the extreme case,  Wigner could expand his description of $A$ to cover the whole lab as an isolated system, including all the lab elements that take part in the measurement, the environment that may dissipate information, and all degrees of freedom that can become correlated with the outcome. In this case, Wigner may be able to describe the final global state as an  entangled state, like for instance $\frac{1}{\sqrt{2}}(\ket{00}_{RA}+\ket{11}_{RA})$.  
While Alice has the subjective experience of observing a definite outcome, Wigner describes instead an entangled state between her memory, the system measured, and the rest of her lab and environment, which is sometimes summarized as  `Wigner models Alice in a superposition of  having seen $0$ and $1$'. We will later explore more complex settings where agents apply quantum theory according to different interpretations; for now the key message is that in order to describe a  measurement performed by some agent, we must specify from whose perspective we want to model it.

\begin{figure}[t]
\centering

\begin{subfigure}{0.4\textwidth}
    \includegraphics[scale=0.2]{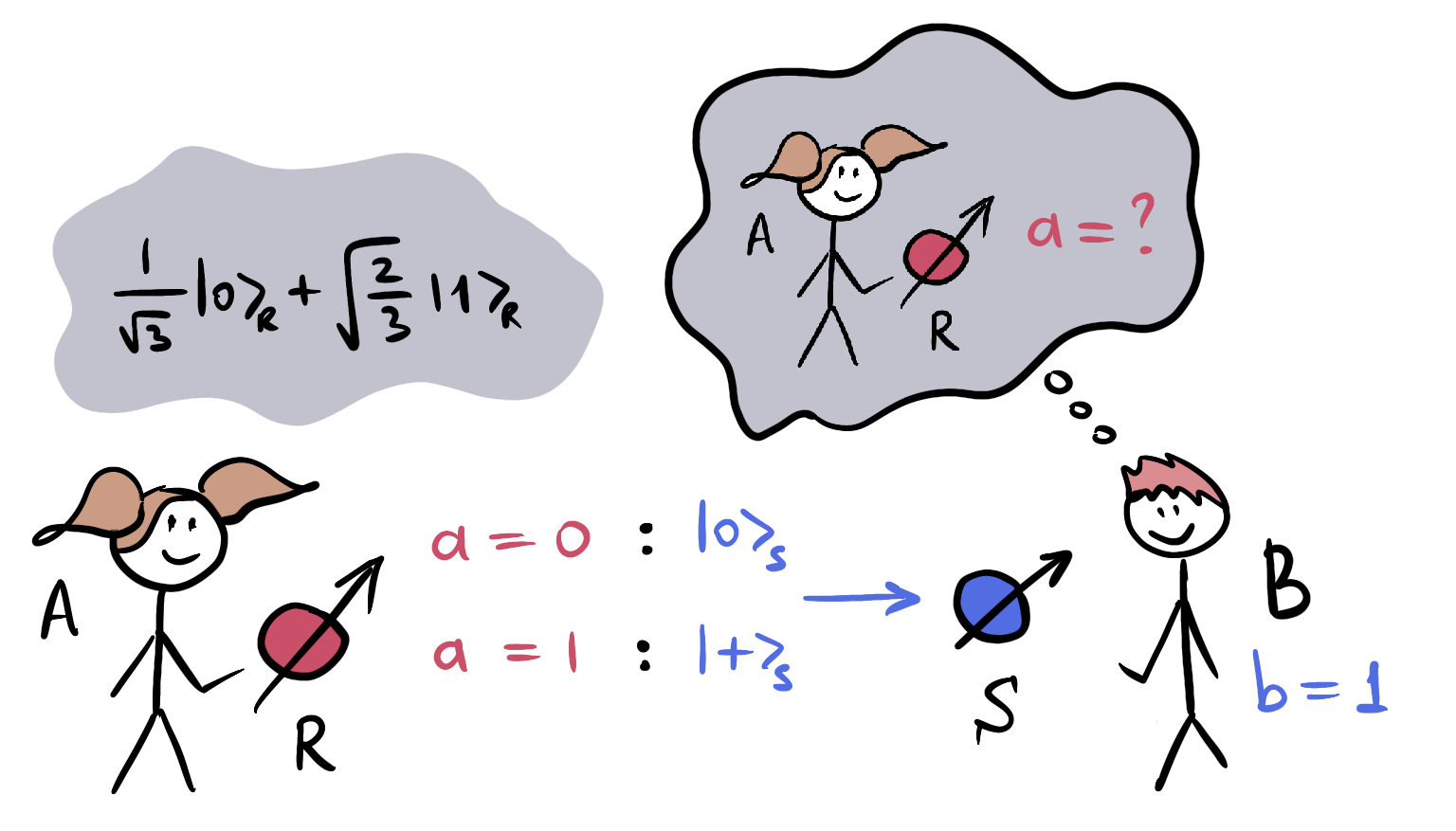}
    \caption{{Experimental setup.} }
        \label{fig:bob-pic}
\end{subfigure}
\qquad \qquad
\begin{subfigure}{0.4\textwidth}
    \Qcircuit @C=1.3em @R=1.3em {
        \push{\hspace{7em}} &\lstick{ \sqrt{\frac13}\ \ket 0_R + \sqrt{\frac23}\  \ket1_R} & \ctrl{1} &  \qw & \qw &  \qw \\
        &\lstick{\textbf{Alice's outcome: } \ket 0_A} & \targ & \ctrl{1} &  \qw & \qw \\
        &\lstick{\textbf{Qubit $S$: } \ket 0_S}& \qw & \gate{H} &   \ctrl{1} & \qw  \\
        &\lstick{\textbf{Bob's outcome: } \ket 0_B} &  \qw & \qw &  \targ & \qw \\
    }
    \caption{Coherent view of the measurements and preparations in the experiment.}
    \label{fig:fr-prep-circuit}
\end{subfigure}
    \caption{{ \bf  A simple quantum inference setup.} Alice measures a qubit $R$, and prepares another qubit $S$ in state $\ket0_R$  or $\ket+_R$ depending on her measurement outcome. Bob then measures $S$, obtains result $b\in \{0,1\}$, and has to reason about Alice's outcome.  On the right is a circuit version of the experiment (up to the reasoning phase), from the perspective of an external agent Wigner who models both Alice and Bob as coherent quantum systems with quantum memories $A$ and $B$. Incidentally, this is the first step of the Frauchiger-Renner experiment. }
\end{figure}

\subsection{Logical reasoning as quantum circuits}

\paragraph{Circuits implement abstract logic.} A summary of our approach is: initial beliefs of agents are encoded in the initial states of their brains, and belief update, as governed by abstract logic rules, is encoded as fixed unitary evolutions on their brains, so that the final brain state encodes their updated beliefs. This unitary applies the logical machinery to arbitrary initial beliefs, much like a program updates the state of a classical memory (or a tape in a Turing machine).

\paragraph{A simple experimental setting.} To illustrate how we model reasoning, we introduce a simple example of two agents (Figure~\ref{fig:bob-pic}), where Bob has to reason about Alice's outcome. Later we will expand this example. 
\begin{itemize}
\item [$t=1$] Alice measures a qubit $R$, initially in state $\frac{1}{\sqrt 3} \ket 0_R + \sqrt{\frac{2}{3}} \ket 1_R$, in basis $\{\ket{0}_R,\ket{1}_R\}$, and records the result $a$ in her memory.  
\item [$t=2$] Alice prepares a qubit $S$ depending on her outcome: $\ket 0_S$ for outcome $a= 0$, and $\frac{1}{\sqrt 2}(\ket 0_S + \ket 1_S)$ for  $a=1$. She gives this qubit to Bob.
\item [$t=3$] Bob measures system $S$ in basis $\{\ket{0}_S,\ket{1}_S\}$, and records the result in his memory.
\item [$t=4$] Bob tries to guess what Alice's outcome was at $t=1$. He can say either ``I am sure that $a=0$'', ``I am sure that $a=1$'' or ``I cannot guess her outcome with certainty.''
\end{itemize}

\paragraph{Abstract reasoning steps.} First let's see how we would reason in abstract about this experiment. Suppose that Bob obtains outcome $b=1$. What does that tell him about Alice's outcome?  Two possible inferences are $b=1 \Rightarrow a=0$ and $b=1 \Rightarrow a=1$,
In this case, it is easy to determine which inference holds deterministically: if Alice had obtained outcome $0$, she would have prepared system $S$ in state $\ket 0_S$, and Bob could not have measured $1$. Hence, the initial state must have been $\ket+_S$, which means that Alice had obtained outcome $1$, and Bob can make a certain inference $b=1 \Rightarrow a=1$. On the other hand, if Bob obtains outcome $b=0$, he cannot make a deterministic retrodiction, as both options for the initial state $\ket0_S$ and $\ket+_S$ are compatible with his outcome.

\begin{figure}[t]
  \Qcircuit @C=1.3em @R=1.3em {
     && \mbox{\blue{measurement}}  
        \gategroup{2}{3}{3}{3}{.7em}{^\}}
        \gategroup{3}{4}{9}{7}{.7em}{--} \\
    \push{\hspace{20em}} 
    &\lstick{\textbf{qubit measured: } \ket \psi_S} 
        & \ctrl{1} & \qw & \qw &  \qw & \qw & \qw \\
    &\lstick{\textbf{Bob's outcome: } \ket 0_B} 
        & \targ & \ctrlo{1} & \ctrl{2}  & \ctrlo{3} & \ctrl{4} & \ustick{\green{  U_{\text{Bob}}  }}  \qw\\
    &\lstick{\textbf{inference: } [b=0 \Rightarrow a=0] \quad \red{\ket 0}_{I^{0, 0}}} 
        & \qw & \ctrl{4} & \qw &  \qw & \qw & \ustick{\green{  \mbox{\qquad \qquad (reasoning)} }}  \qw \\
    &\lstick{\textbf{inference: } [b=1 \Rightarrow a=0] \quad \red{\ket 0}_{I^{1, 0}}} 
        & \qw & \qw & \ctrl{3} &  \qw & \qw & \qw \\
    &\lstick{\textbf{inference: } [b=0 \Rightarrow a=1] \quad \red{\ket 0}_{I^{0, 1}}} 
        & \qw & \qw & \qw &  \ctrl{3} & \qw & \qw\\
    &\lstick{\textbf{inference: } [b=1 \Rightarrow a=1] \quad \green{\ket 1}_{I^{1, 1}}} 
        & \qw & \qw & \qw & \qw  & \ctrl{2} & \qw\\
    &\lstick{\textbf{Bob's prediction: } a=0 \quad \ket 0_{P^0} }
        & \qw & \targ & \targ &  \qw & \qw & \qw \\
    &\lstick{\textbf{Bob's prediction: } a=1 \quad \ket 0_{P^1}} 
        & \qw &  \qw & \qw & \targ & \targ & \qw
    }
    \caption{{ \bf Modelling agents as coherent reasoning circuits.}   Our physical model of Bob, as a quantum  reasoning agent, has three main parts: a \textit{memory qubit} $B$, where he stores the measurement outcome;  \textit{inference qubits} $\{I^{b,a}\}_{b,a}$, which encodes the agent's initial beliefs about the validity of different inferences; \textit{prediction} qubits $\{P^a\}_a$ that express his  conclusions regarding Alice's outcome. Each inference qubit $I^{b,a}$ is initialized in state $\ket1_{I^{b,a}}$ if the inference  $b \Rightarrow a$ holds deterministically in this setting (from Bob's perspective, as explained later), and $\ket0_{I^{b,a}}$ otherwise. Bob's measurement is modelled (by an external agent) as a CNOT gate between the system measured and Bob's outcome register.  After performing the measurement, the reasoning is modeled as a series of doubly-controlled gates that depend on the outcome and inference qubit registers. They affect the state of the prediction qubit: a final state of $\ket1_{P^a}$ means ``Bob guesses that Alice's outcome was $a$ deterministically.'' For the controlled gates, note that $\circ = X \bullet X$. In this case, and given the way the inferences are initialized, only the last doubly-controlled gate (controlled on $I^{1,1}$) acts non-trivially in practice. We denote Bob's reasoning unitary by circuit $U_{\text{Bob}}$ (dashed region).}
    \label{fig:bob-circuit-full}
\end{figure}
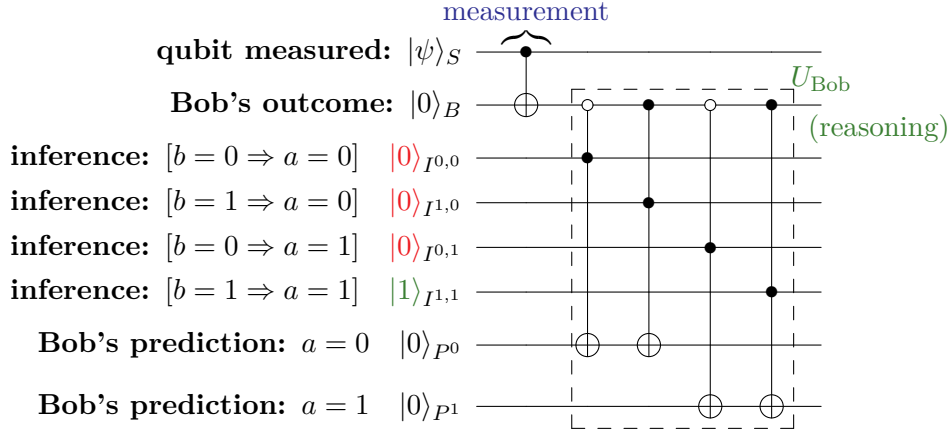

\paragraph{Physical encoding of preparations and measurements.} From the perspective of an external agent (like Wigner) who models Alice and Bob as coherent quantum systems interacting with qubits $R$ and $S$, the first three steps of the experiment can be modelled as a simple quantum circuit (Figure~\ref{fig:fr-prep-circuit}), where the measurements are CNOTs, and Alice's conditional preparation of $S$ is a Hadamard gate controlled on her memory register.  It remains  to model the last part of the experiment: Bob's reasoning.

\paragraph{Physical encoding of inferences and reasoning.}  Now we describe how an external agent (like Wigner, who sees Bob as an evolving quantum system) could represent Bob's reasoning process as a quantum circuit.\footnote{How Bob models his own evolution depends on his interpretation of quantum theory; more on this in the next section.} First we expand our model of Bob's brain to include more quantum registers and gates (Figure~\ref{fig:bob-circuit-full}).  In our model, Bob's prior beliefs about the truth value of prepositions (like ``if I observe $b=1$ I can conclude that $a=1$'' or ``smoke implies fire'') are encoded in the initial state of some qubits in his brain, and can change depending on the experimental setting. On the other hand, the abstract logical system used by Bob to process his beliefs and observations is implemented by the unitary circuit. For example, a series of controlled gates implement the knowledge distribution axiom, $(B \wedge (B \Rightarrow A) ) \Rightarrow A$, and allow Bob to make predictions. We will now explain this model in more detail.
We dedicate an \textit{inference qubit} $I^{b,a}$ to each possible inference $b\Rightarrow  a$ available to an agent: the qubit is initialized in  state $\ket 1_{I^{b,a}}$ if Bob believes that the inference  holds, and $\ket 0_{I^{b,a}}$ if it doesn't. We will explain in Section~\ref{sec:logic-inter} how to model Bob's decision process to find the truth value of possible inferences and initialize these qubits; for now let us skip this step and assume that Bob already  concluded that
 according to the Born rule only the inference $b=1 \Rightarrow  a=1$ is true.  We model this by initializing the corresponding qubit in  state $\ket 1_{I^{1,1}}$, and leaving the other inference qubits in state $\ket 0_{I^{b,a}}$. 
The missing piece now is how Bob reaches a prediction about Alice's outcome: he must apply some logical reasoning to process the information present in his list of true inferences and his measurement outcome register. In other words, we are looking for a quantum version of applying the logical reasoning step ``I observed $b$'' $\wedge$ ``I know that $b\Rightarrow  a$ '' $ \Rightarrow $ ``I conclude that $a$ holds'', checking for all possible combinations of $a$ and $b$.
For this we equip Bob with \textit{prediction qubits} $\{P^a\}_a$, one for each possibility for Alice's outcome $a$: a final state of $\ket1_{P^a}$ is interpreted as ``Bob guesses that Alice's outcome was $a$ with certainty''. These qubits are initialized to $\{\ket 0_{P^a}\}_a $ (which corresponds to ``Bob cannot guess that Alice's outcome was $a$ with certainty''). He applies this logical reasoning  by  performing sequential updates of prediction qubit, each modelled as a Toffoli gate (a doubly-controlled gate)  conditioned on his  outcome register and the corresponding  inference qubit.  

\paragraph{A note on complexity and robustness.}  We do not claim that this is the most compact way to model reasoning processes; it is nevertheless an approach that generalizes well to more complex settings, for example including more agents, each with a larger brain, processing  more inferences, observations and predictions.  One advantage is that it is easier to then customize which inferences are valid as a result of applying a different physical theory, interpretation, or logical rule system. The actual reasoning circuit (whose gates represent how agents process inferences) stays the same, and we only need to initialize the inference qubits appropriately. This is akin to saying ``the logical structure used by each agent is stable, but they can instantiate each premise as true or false.''

\begin{figure}[t]
    \Qcircuit @C=1.3em @R=1.3em {   
        \push{\hspace{16em}} 
        &\lstick{\textbf{Qubit $R$: } \sqrt{\frac13}\ \ket 0_R + \sqrt{\frac23}\  \ket1_R}     
            & \ctrl{1}  &  \qw & 
            \ustick{\mbox{ \blue{Alice's actions}}} 
        \qw &  \qw & \qw & \qw & \qw & \qw & 
        \rstick{\mbox{\blue{Check}}} 
        \\
        &\lstick{\textbf{Alice's outcome: } \ket 0_A} 
            & \targ & \ctrl{1} &  \qw & \qw  & \dstick{\mbox{\blue{Bob's actions}}}  
            \qw &   \qw &\qw & \qw & \meter& \rstick{\blue{a}} \cw &  \\
       &\lstick{\textbf{Qubit $S$: } \ket 0_S}
            & \qw & \gate{H} &   \ctrl{1} & \qw &  \qw & \qw &  \qw & \qw   & & & & &  \\
        &\lstick{\textbf{Bob's outcome: } \ket 0_B} 
            &  \qw & \qw &  \targ & \ctrlo{1} & \ctrl{2} &  \ctrlo{3} & \ctrl{4} & \qw & & & \\
        &\lstick{\textbf{inference: } [b=0 \Rightarrow a=0] \quad \red{\ket 0}_{I^{0, 0}}}
            &  \qw & \qw & \qw & \ctrl{4} & \qw & \qw &  \qw & \qw & & &  \\
       &\lstick{\textbf{inference: } [b=1 \Rightarrow a=0] \quad \red{\ket 0}_{I^{1, 0}}}     
        &  \qw & \qw & \qw & \qw & \ctrl{3} &  \qw & \qw & \qw & & &  \\
        \push{\hspace{11em}} &\lstick{\textbf{inference: } [b=0 \Rightarrow a=1] \quad \red{\ket 0}_{I^{0, 1}}}     &  \qw & \qw & \qw & \qw  & \qw & \ctrl{3} & \qw & \qw & & & \\
       &\lstick{\textbf{inference: } [b=1 \Rightarrow a=1] \quad \green{\ket 1}_{I^{1, 1}}} 
            &  \qw & \qw &  \qw & \qw & \qw & \qw & \ctrl{2} & \qw & & & & \\
       &\lstick{\textbf{Bob's prediction: } a=0 \quad \ket 0_{P^0} }
            &  \qw & \qw & \qw & \targ & \targ &  \qw & \qw & \qw & \meter& \rstick{\blue{p^1}} \cw  & \\
       &\lstick{\textbf{Bob's prediction: } a=1 \quad \ket 0_{P^1}}
            &  \qw & \qw & \qw & \qw &  \qw & \targ & \targ & \qw & \meter & \rstick{\blue{p^0}} \cw & \push{\ }
      \gategroup{1}{3}{3}{4}{.7em}{--} 
      \gategroup{3}{5}{10}{9}{.7em}{--} 
      \gategroup{2}{11}{10}{13}{.7em}{--}
    }
    \caption{{ \bf Simulation of the whole experiment.}   An external Wigner can now simulate the whole experiment as a unitary circuit evolution, from Alice's measurement and conditional state preparation to Bob's measurement and reasoning. To check the validity of Bob's predictions, Wigner can measure the registers $A$, $P^0$ and $P^1$, and see if the classical outcomes are compatible.}
    \label{fig:alice-bob-circuit-fr1}
\end{figure}
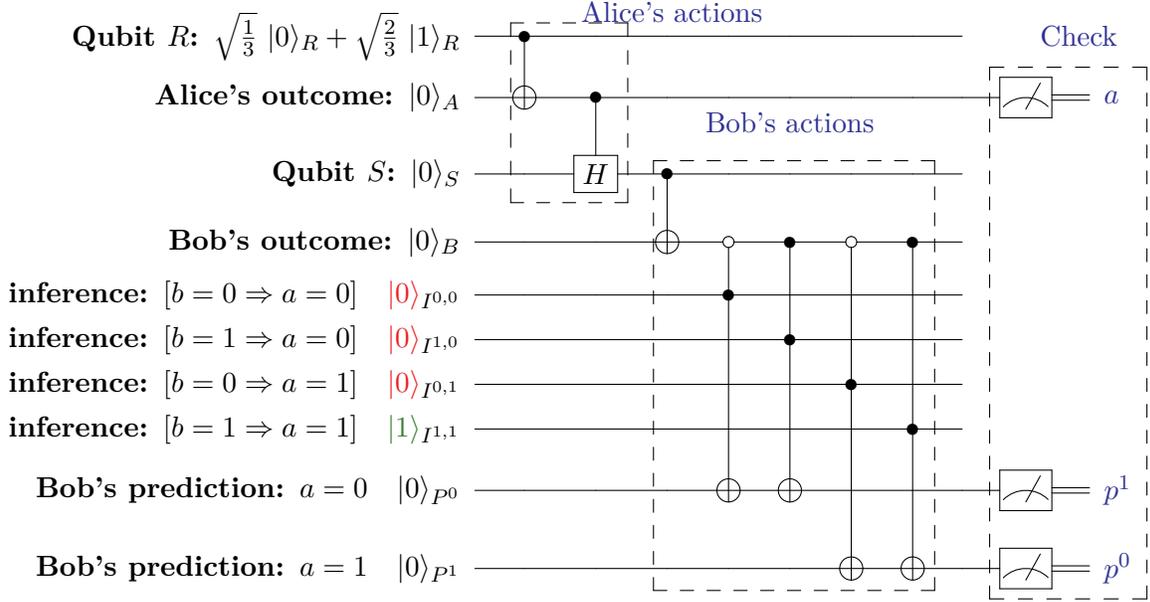

\paragraph{Simulating the experiment and consistency checks.} From the external agent's perspective, the simulation of the whole experiment consists of composing the circuits of the two previous steps: Alice's coherent measurement and state preparation of Figure~\ref{fig:fr-prep-circuit}, and Bob's  coherent measurement and reasoning of Figure~\ref{fig:bob-circuit-full}. The result is in Figure~\ref{fig:alice-bob-circuit-fr1}.  One way to check  the consistency of Bob's reasoning is to measure his two prediction registers as well as Alice's memory at the end, and see if the classical outcomes are compatible. For example, we can repeat the experiment many times and check that whenever Wigner measures on Bob's prediction register $p^1=1$, he also finds that the outcome of measuring Alice's memory is indeed  $a=1$. (For a more quantum flavour, Wigner could leave the circuit in a quantum state, and study quantum correlation measures between Bob's prediction qubits and Alice's memory.) The only missing step is deciding how Bob initializes his inference qubits. 

%% file: logic-inter.tex
\paragraph{Theories applied by agents.} In our framework, each agent can decide which inferences are valid, and initialize their inference qubits accordingly. To do this, agents apply a given theory to the experimental setting. We employ \emph{theory} as a broad concept that includes the version and interpretation of quantum mechanics they apply, their assumptions about other agents, the logical axioms they follow, and the information available to them about the experimental setting. Agents can decide on the overall theory they apply before running the experiment. As such, these considerations are processed before the experiment to obtain the list of valid inference rules, as we will explain ahead. Their direct impact on the experiment is then simply the initialization of the inference qubits to different values.

\subsection{Interpretations of quantum theory and inferences}

\paragraph{Defining interpretations.} An interpretation of quantum theory determines what agents can infer from experimental outcomes (for example ``the global state has now collapsed'', ``I know that it has not collapsed, but I also know that Alice thinks it has'', ``I am actually now entangled with the system measured, and soon Bob will be entangled too'').  
In our software, a minimal description of an interpretation defines two functions, which specify how to perform forward and backward inference procedures. For each quantum measurement $\mathcal M_B$ performed by an agent (say Bob) in the course of the experiment, forward inference will compute what he can conclude about events that will happen after  $\mathcal M_B$   (depending on his measurement result), and the backward inference computes retrodictions about events that happened before  $\mathcal M_B$. 

\paragraph{Inference mechanisms.} In practice, this is applied to concrete experiments in the following way: the protocol of the experiment specifies which inferences we want to check (for example  ``Bob must reason about the different options for $\{b \implies a\}_{a,b}$'' in the previous example). For each of these inferences, the program takes the interpretation used by the agent and automatically runs  quantum simulations of the relevant part of the experiment, from the perspective of the agent, and according to their interpretation, to see if there are simulations compatible with the inference $b \implies a$; from here it derives  the list of inferences that the agent would consider valid. Which ``portion of the experiment'' is relevant for the simulation is derived from the forward and backward inference rules specified by each interpretation. For example, a many-worlds interpretation may ask us to simulate and keep track of the global state evolution including several agents, while a collapse interpretation can be more economical on what it needs to reach predictions.
In summary, given a particular protocol, the software goes through the following steps to simulate it:
\begin{itemize}
   \item[1.]
    Define each step of the protocol by specifying the action (e.g.\ measurement, preparation, reasoning) and the domain of the action (agents and systems involved in the action).
    
    \item[2.] Determine how many qubits to allocate to each agent (depending on the complexity of the predictions they have to make) and other systems. 
    
    \item[3.]
     Initialize the inference qubits. For this, we run the function making the inference for each agent, and fill in classical inference tables. 
     To apply the inference function, the reasoning agent  fixes some of the outcomes and runs a restricted quantum simulation of some steps of the experiment, in order to check the validity of possible inferences; they repeat this for every possible outcome combination. We explain in more detail how these processes are modeled in the examples ahead. 
    
   \item[4.]
    Initialize the inference qubits of all agents, run the protocol as a whole, and output the predictions of all agents.

\end{itemize}

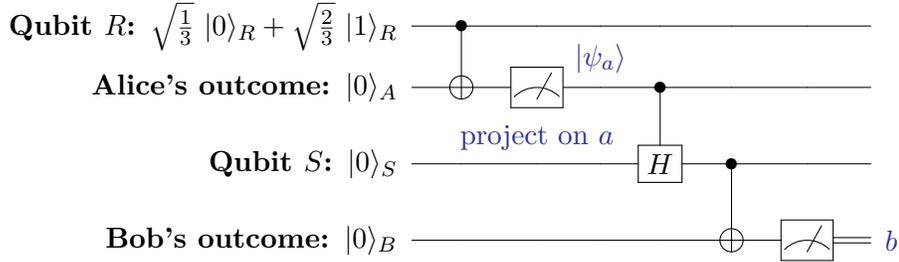
\begin{figure}[t]

    \Qcircuit @C=1.3em @R=1.3em {    
        \push{\hspace{16em}} 
        &\lstick{\textbf{Qubit $R$: } \sqrt{\frac13}\ \ket 0_R + \sqrt{\frac23}\  \ket1_R}     & \ctrl{1} &  \qw & \qw &  \qw & \qw & \qw & \qw   \\
        &\lstick{\textbf{Alice's outcome: } \ket 0_A} 
            & \targ  & \meter&   \ustick{ \blue{\ket{\psi_a}}} \qw &   \ctrl{1}  &  \qw & \qw &  \qw \\
        &\lstick{\textbf{Qubit $S$: } \ket 0_S}
            & \qw& \ustick{\mbox{\blue{project on $a$}}}\qw &\qw   & \gate{H}  & \ctrl{1} & \qw & \qw & \\
        &\lstick{\textbf{Bob's outcome: } \ket 0_B} 
            & \qw & \qw & \qw &\qw  &  \targ & \meter & \rstick{\blue{b}} \cw &  \\
    }
    \caption{{ \bf Bob's restricted quantum simulation.} In order to determine the validity of the various possible inferences $b\implies a$, Bob runs a restricted quantum  simulation of the experiment. As he needs to reason about Alice's outcomes, he simulates a measurement on her memory, and proceeds with independent simulations with the post-measurement states corresponding to each of the outcomes $a$. After his measurement at the end, he compares the likelihood of obtaining $b\wedge a$, and classically determines whether $b\implies a$ applies.  }
    \label{fig:bob-restricted-simulation}

\end{figure}

\paragraph{Example: Bob's restricted quantum simulation.} Still following our previous example, let us see how Bob can make a restricted (quantum) simulation to initialize his inference qubits about Alice's outcome (Figure~\ref{fig:bob-restricted-simulation}). In this simple case the exact interpretation of quantum mechanics is not yet critical. Bob models Alice's lab's evolution, and projects the state to one of Alice's memory subspaces, corresponding to her possible outcomes $a= 0, 1$.  
\begin{itemize}
    \item In Bob's simulation  $R$ is initialized in $\frac{1}{\sqrt 3} \ket 0_R + \sqrt{\frac{2}{3}} \ket 1_R$. Simulated Alice measures $R$ and writes the result down to her memory, through a CNOT gate,
        \begin{gather*}
           \frac{1}{\sqrt 3} \ket 0_R\ket 0_A + \sqrt{\frac{2}{3}} \ket 1_R \ket 1_A. 
        \end{gather*}
    Now Bob simulates a measurement in Alice's memory, and projects the post-measurement state onto the subspaces corresponding to the different outcomes $a=0,1$ independently.  His simulation carries on further for each of Alice's outcome options separately until Bob's measurement, where Bob also models his own memory as a qubit.
    
    \item Case $a=0$: Bob continues with the post-measurement state corresponding to $a=0$, $\ket 0_R\ket 0_A$.
    \begin{enumerate}
        
        \item Simulated Alice prepares  $S$ in state $\ket0$ and sends it to simulated Bob, who measures it, coherently copying the result to his memory,
        \begin{gather*}
            \ket 0_R\ket 0_A \ket 0_S \ket 0_B.
        \end{gather*}
        
        \item Bob can now measure simulated Bob's memory. From this case Bob can conclude that $a=0 \implies b=0$. To check the converse he must still inspect the other case.
    \end{enumerate}
    
    \item Case $a=1$: Bob continues with the post-measurement state corresponding to $a=1$, $\ket 1_R\ket 1_A$.
    \begin{enumerate}
        
        \item Simulated Alice prepares qubit $S$ in state $\ket+_S$ and sends it to simulated Bob, who measures it,
        \begin{gather*}
            \frac{1}{\sqrt 2}\ket 1_R\ket 1_A \ket 0_S \ket 0_B + \frac{1}{\sqrt 2}\ket 1_R\ket 1_A \ket 1_S \ket 1_B.
        \end{gather*}
        
        \item Now Bob measures his simulated self's memory, and concludes that $a=1 \implies (b=0 \vee b=1)$.  In other words, if he obtains $b=0$, Alice could have measured $a=1$; if he gets $b=1$, Alice could have measured $a=1$ as well.
    \end{enumerate}
    \item  Finally, Bob's inference table $\{b\implies a\}_{a,b}$ is constructed by logically combining the analyses of both cases. 
    Bob concludes that the only certain inference is $b=1 \Rightarrow a=1$, so that inference qubit will be initialized as $\ket{1}_{I^{1,1}}$, and the others as $\ket0_{I^{a,b}}$.  In conclusion, Bob can now initialize his inference qubits as $\ket{\vec x_B}_{I_B} = \ket{0001}_{I_B}$, like we saw in Figure~\ref{fig:bob-circuit-full}.
\end{itemize}

\paragraph{Why running restricted simulations?} In principle, these quantum simulations that agents use to compute inferences could also be compiled in runtime, as a more involved subcircuit in the brains of agents in the global experiment. We chose to split them into restricted pre-computations simply for practical reasons, namely restricted quantum memory space in the NISQ era. Including these quantum simulations as part of the global experiment would require us to expand the models of each agent's brain by a number of qubits that scales badly on the complexity of the protocol (number of agents and experimental steps). In a future where we have access to large-scale quantum computers, it would be interesting to run the full quantum experiment (including these inference steps) in one go. As of late 2022, this does not seem feasible just yet.

\paragraph{Avoiding recursion issues.}
When an agent runs a quantum simulation to make inferences, sometimes they may have to include themselves in the simulation, which in principle could lead to recursion problems. 
Our design avoids recursion issues --- at least in all the cases tested. For direct examples, some of the simulations presented in Appendix~\ref{sec:FR} include a restricted model of the agent running them.   One technique for avoiding recursion is that the restricted quantum simulations include projections of the simulated state into each possible outcome (as the agent is trying to infer what happens when an outcome is observed); this already restricts the degrees of freedom modelled in the simulation, and prevents potential recursion loops. The same can be applied to the agent's own inference qubits: they don't need to be initialized (in the simulation) to the final correct value, but instead the agent can test what happens for different initial values of the inference qubits. In our examples, the simulated inference qubits can be safely left out of the simulations, or initialized in an arbitrary state like $\ket0$.

\paragraph{Currently available interpretations.} As of September 2022, there are two interpretations implemented in the package: neo-Copenhagen and collapse, which we describe briefly here for completeness. In~\cite{Nurgalieva2021}, we discuss how the quantum mechanical thought experiments like FR or Wigner's friend fit into more interpretations, like Bohmian mechanics~\cite{Bohm1952}, QBism~\cite{Fuchs2014}, relational QM~\cite{Rovelli1996, Rovelli1997}, and various versions of many-worlds~\cite{Vaidman2001}. We encourage readers to implement other such interpretations in this package.

\paragraph{Copenhagen interpretation.}
In the Copenhagen interpretation~\cite{Bohr1934,Bohr1958,Heisenberg1958} the modelling of nature is split in two parts: the first part, ``observer'', is the observing system which acquires knowledge by the way of carrying out the experiment, and also includes measuring devices; and the second part is the observed system. The observer's  experiences are expressed in the ordinary (classical) language of physics, while the observed system is described in the language of quantum mechanics. The separation above is called the \textit{Heisenberg cut}~\cite{Atmanspacher1997}. The cut can be understood as being ``objective'', if there is a fundamental property of nature fixing it for all scenarios (conventional Copenhagen), or ``subjective'' -- when the placement of the cut  is determined separately for each observer (neo-Copenhagen~\cite{Brukner2017}). In our implementation, we allow agents to model several other agents as quantum systems, while still retaining classicality themselves. In other words, the Heisenberg cuts of individual agents do not necessarily coincide: Alice can model Bob as a quantum system, while considering herself classical, and Bob can do the reverse.

\paragraph{Objective collapse theories (GRW).}
Objective collapse theories~\cite{Ghirardi1985} are an extension of the existing formalism of quantum theory. They establish dynamics that govern macroscopic and microscopic processes in nature, by adding stochastic and non-linear terms to standard dynamical equations; events of non-unitary collapse depend on phenomenological parameters. The current experimental evidence provides lower and upper bounds on these parameters, which makes collapse theories one of the few falsifiable interpretations in the short term~\cite{Bilardello2016, Bilardello2017,Carlesso2018,Toro2018}. In our implementation, the relevant aspect that we take is that in objective collapse theories only one outcome is observed (with some probability) and the post-measurement state collapses --- this is an objective view shared by all agents who use the interpretation, regardless of who performed the measurement. 
We implement these stochastic collapse theories through a tree class, where an instance of a tree  encapsulates a quantum experiment. The tree's  branches represent what happens after a measurement outcome is observed and the corresponding  probability (much like in settings where classical randomness plays a role).

\subsection{Combining inferences: logic}

\paragraph{Knowledge distribution axiom.} 
A key logical axiom needed for basic reasoning is the \textit{distribution axiom}~\cite{Kripke2007, Fagin2004}.\footnote{More precisely, this is only one of a set of axioms used in multi-agent epistemic settings. We do not review the rest as they are not relevant to the implementation in this paper -- if you want to learn more, consult our modal logic analysis of FR in~\cite{NL2018}.} It allows us to conclude that if an agent knows ``$A \Rightarrow B$'' and ``$B \Rightarrow C$'', then they also know ``$A \Rightarrow C$''. Formally, this is expressed as 
\begin{gather*}
    K_i (A \Rightarrow B) \wedge K_i (B \Rightarrow C) \ \Longrightarrow \ K_i (A \Rightarrow C),
\end{gather*}
where the knowledge operator $K_i$ represents ``agent $i$ knows...''.

\paragraph{Transitivity of knowledge across agents.}
However, to be precise, an agent is not always combining their knowledge directly -- sometimes they reason from the viewpoint of a different agent $j$, which corresponds to stating that ``agent $i$ knows that agent $j$ knows that...''. Then to use the distribution axiom like above, we  require an additional step of trusting the knowledge of another agent as well as their own,
\begin{gather*}
    K_i (A \Rightarrow B) \wedge K_i K_j (B \Rightarrow C) \ \Longrightarrow \ K_i (A \Rightarrow B) \wedge K_i (B \Rightarrow C) \ \Longrightarrow \ K_i (A \Rightarrow C).
\end{gather*}
The condition $K_i K_j \phi \ \Longrightarrow K_i \phi$ holds trivially for classical multi-agent settings where all agents are rational and have access to the same set of common knowledge (for example, ``we all assume the Born rule'',  ``the experimental protocol follows these steps'', ``each agent only sees one outcome of a measurement'', etc). However, in quantum scenarios this condition might not necessarily hold, for example, due to the duality of perspectives we have seen in Wigner's friend experiment. Wigner may not want to adopt the knowledge of Alice, which he sees as more restricted: Alice doesn't consider that she is entangled with the system measured. Hence, agents in quantum settings require additional constraints, which  governs the way their knowledge can be combined. In~\cite{NL2018, Vilasini2019}, we called the set of these conditions \textit{trust relations}.\footnote{In retrospect, `trust' was an unfortunate term. We now think of it as a relation that tells us `Alice believes that she has a model for Bob's reasoning, and tries to reason with that model from his perspective'.} Conceptually, these conditions are related to the original \textbf{C} assumption in the FR paradox~\cite{Frauchiger2018}, which postulates that agents reason from each other's viewpoints. Trust relations can restrict and disallow such reasoning for particular set of agents. 

\paragraph{Trust structures.}
The trust structure currently implemented is the trivial one, where agents trust each other irrespective of the actions they perform on each other. That is, the Consistency module allows agents to combine the inferences of any two agents: it takes as arguments any two inference tables encoding $A\implies B$ and $B\implies C$, and returns the inference table with the conclusions $B \implies C$. This can be refined to restricted trust structures, for example by checking which users provided the original inferences and whether they are allowed to combine them. 
As an intuitive example, we could impose that only agents with commuting measurements could trust each other. 
Yet this restriction may not be sufficient to avoid paradoxes:  in the FR thought experiment with the neo-Copenhagen interpretation, combining knowledge across commuting agents leads to a contradiction (even when they don't employ that trust structure fully and only reason about adjacent agents). See Appendix~\ref{sec:FR} for a walk-through of the experiment.

%% file: conclusions.tex
\subsection{Applications and insights}

\paragraph{Application: simple experimental settings.}  For examples of simple examples of experimental settings, see Appendix~\ref{appendix:examples}; all experiments are also available as open Jupyter notebooks.  These are useful as litmus tests of new proposals to avoid paradoxes. 
That is, suppose that the user wants to avoid contradictions in complex thought experiments like FR, and tries to restrict the logical axioms or to apply a bespoke interpretation (for example by disallowing any trust chains, or implementing an extreme version of many worlds). It is not sufficient to show that those settings avoid the paradox: they should also allow agents to make standard inferences in very simple, everyday setups (like Alice and Bob measure each half of a Bell state, and should predict each other's outcomes). The user can then test their interpretation and logical axioms in this kind of scenarios --- and indeed often those restrictions don't allow Alice and Bob to make the simple predictions of this example \cite{NL2018}.

\paragraph{Application: clarifying the FR experiment.} 
In Appendix~\ref{sec:FR} we model the whole FR experiment and explore different variations on how agents choose to reason about each other. 
 A common criticism of the FR thought experiment is that Alice's prediction of $w=\text{fail}$ is undone by another agent's actions, and cannot be accessed after Wigner's measurement, so there isn't a direct contradiction.  However, our implementation shows that a physical trace of Alice's prediction (through all the chains of reasoning) actually survives until the end of the experiment in the physical brain of another agent (Ursula), and can be observed even by Wigner at the same time as his actual outcome $w=\text{ok}$.  We are not aware of examples of this kind of live multi-agent contradictions in classical physics (and non-contextual theories in general). It is up to the community to decide if this is problematic, and how to address it.

\paragraph{Foundational outlook: a platform for testing approaches.}
The package introduced here allows us to build simple models of agents and their logical interactions in the context of thought experiments in quantum settings. This includes a modular set of the priors that agents use to derive their conclusions, which depends heavily on the interpretation of quantum mechanics that they apply. 
For many decades, interpretations of quantum mechanics have been considered a somewhat philosophical and abstract discussion. Implementing them in  our  software in a mathematically rigorous way and testing them in different thought experiments brings the philosophical disagreement to a technical level, and provides additional motivation and tools for formalizing existing approaches.

\paragraph{Computational outlook: a test network of quantum computers.}
The transitivity of reasoning discussed here mimics the behaviour of information across components of a large quantum computer, where different parts of the machine might measure each other in a complicated sequence, and still need to retain consistency of information stored in their respective quantum memories. Our framework can serve as a good testing ground for the  logical systems that can be implemented in large-scale quantum devices.

\subsection{Directions for future work}

\paragraph{Trust structure compiler.} A useful  extension would be a ``trust compiler'', that is a feature to automatically compute and implement the trust relation structures of an experiment (from the input `protocol', `interpretations for each agent' and `general user-defined conditions for trust'). We are cautiously optimistic about the ease of implementation of this feature: we can write a beta program that works well for familiar sets of conditions and experiments, but, like most things in quantum foundations, it's hard for us to predict whether it would be general enough to handle unexpected rules. What if a user would like to test consistency of an interpretation of quantum mechanics under a convoluted condition like ``I cannot trust the knowledge of agents who will be measured in this precise basis, in this time interval''? (The answer in this case is that it seems awfully fine-tuned to a given experiment and would probably not hold up in slightly modified thought experiments, and that maybe the user should program it themselves.) It could also be a more sensible and general condition for trust in quantum settings, but of a structure that we cannot yet predict,\footnote{Given that most of our research focuses on finding such conditions, we'd venture that we are relatively well-positioned to guess that it must be surprising.} and would not fit our trust compiler's output requirements. 

\paragraph{Logical axioms compiler.}
More generally we should be able to take in a selection of arbitrary logical axioms specified by the user~\cite{Hintikka1962, Hilbert1999}, and process them (together with the experimental setting and agents' specifications) into a final trust structure and set of instructions for how each agent makes their predictions. The problem here is again generality: the first step is to identify a framework general enough to cover many types of logical axioms, and standardized enough that it could be coded into our ``logical axioms compiler''. A good starting place is consulting with the classical computational logic community~\cite{Kripke2007, DELStanford, Ditmarsch2007} --- however, we suspect that once again there could be suggestions for new quantum axioms that wouldn't fit a previously standard specification. Once again, the solution is to generalize our base framework, when such promising axioms appear.

\paragraph{Probabilistic predictions and quantum logic.} In the applications studied we focused on deterministic predictions. In principle the software has all the pieces in place to handle probabilistic predictions, and we believe it would be straightforward to implement a judgement of contradictions for finite probabilities. It should also be suitable to process and generalize some types of quantum logic~\cite{QLStanford, Baltag2011, Chiara2013part1, Chiara2013part2}, as the agents' memories, inferences and predictions are already encoded as quantum states. We'll welcome these extensions when there is sufficient interest.

\paragraph{More compact models for agents.} We would like to find  more  compact circuit models of agents; these should still be robust and modular in complex settings, but not require as many qubits and gates to run. Stating from a more economical basic agent model would facilitate later expanding it to more complex information-processing systems.  One simple option would be to allow for $d$-dimensional quantum registers and not just qubits, which would allow us to use a smaller Hilbert space over all fewer multiplexed gates. For example, suppose that Bob's measurement has a total of $d$ outcomes (coherently copied to his $d$-dimension outcome register), and he has to make a prediction about Alice's $k$ possible outcomes. The possible inferences could  be encoded in $d$ inference registers $\{I_b\}$ of dimension $k+1$ each. Each of the registers' initial state $\ket{a}_{I_b}$ would encode ``when Bob sees outcome $b$ he predicts that Alice obtains $a$'', with $\ket{a=0}_{I_b}$ to represent ``In this case Bob cannot make a sure prediction about Alice''.  The total dimension of the inference registers would then be $d^{k+1}$ as opposed to $2^{dk}$ of our original scheme. The unitary reasoning could be implemented by $d$ doubly multiplexed gates, controlled on the outcome and inference registers, and targeting prediction registers. We have not yet tested the implications of this approach, for reasons detailed ahead.

\paragraph{Libraries for quantum registers of arbitrary dimension.} A technical implementation issue  for the above proposal is that ProjectQ, and most quantum programming languages, don't allow us to implement registers and gates of dimensions that are not of the form $d=2^n$ --- that is that don't correspond to a natural number $n$ of qubits. One reason is that quantum programming languages want the compilation of any program to be compatible with standard quantum hardware implementations, which require qubit `assembly' instructions to run. Therefore we would have to write an extra library for ProjectQ to translate code written for $d$ dimensions to the smallest number of qubits necessary, which is a new research project on its own.  With our current knowledge we cannot guess if the compilation of the final program, including this translation, would in the end be more computationally efficient than our current model.

\paragraph{More interpretations.}
We look forward to implementing more interpretations of quantum theory:  this would help us test not only the interpretations but also the generality and modularity of the software. This is an open-source project and community input is very much appreciated, in particular for this point --- we may have a good insight into how Quanundrum works, but only the reader knows the details of their personal interpretation of quantum theory. 

\paragraph{Dynamic experiments and quantum reference frames.}
In the long term we would like to expand the Protocol model to cover dynamic experimental settings, where the order of operations can (coherently) depend on previous outcomes. Another ambitious, long-term generalization would be to cover cases where agents can have different quantum reference frames (both abstract frames and concrete quantum systems employed as references). This would be a step towards simulating relativistic quantum thought experiments. As stepping stone, classical special relativity should be easier to integrate in the package: it would require us to rethink the applicability of the current ``forward and backward inference'' functions, and to generalize them for predictions between any two points in an arbitrary causal order structure.

\paragraph{Beyond quantum theory.} All interpretations of quantum theory define in fact new physical theories, which coincide in some regimes --- typically, those that are experimentally accessible. For example, in most interpretations, agents expect their measurement statistics to follow the Born rule, and model the evolution of small isolated systems as unitary transformations on a Hilbert space. In most of them, they can even model other agents as quantum systems evolving in an environment; the differences between different versions of quantum theory arise when agents consider what happens globally when they perform measurements, the dynamics at very large scales,  what hidden variables the theory follows or even how it behaves in relativistic regimes. Our software package can handle these differences; but what happens for completely different theories, without the common quantum basis? If we would like to simulate agents in Generalized Probabilistic Theories \cite{Vilasini2019}, for example, we need first a model for dynamics in the target theory, so that we could implement an analog of the circuit version of each agent's brain, according to the principles of that theory. For this, a quantum programming language like ProjectQ may not be the most appropriate, although most of the modular structure of Quanundrum and main components could be easily adapted.

%% file: appendix_examples.tex
 Here we review  examples of simple thought experiments where the reasoning of agents can be modelled by small quantum simulations. These simple examples can be used as canaries: if a model (with a physical theory, interpretation and set of logic rules) leads to contradictory conclusions by agents, this suggests that the model has limited predictive power and is not widely applicable. Links to pedagogical Jupyter notebooks are provided.

\begin{figure}[H]
    \centering
    \begin{subfigure}{0.45\textwidth}
    \centering
    \includegraphics[scale=0.3]{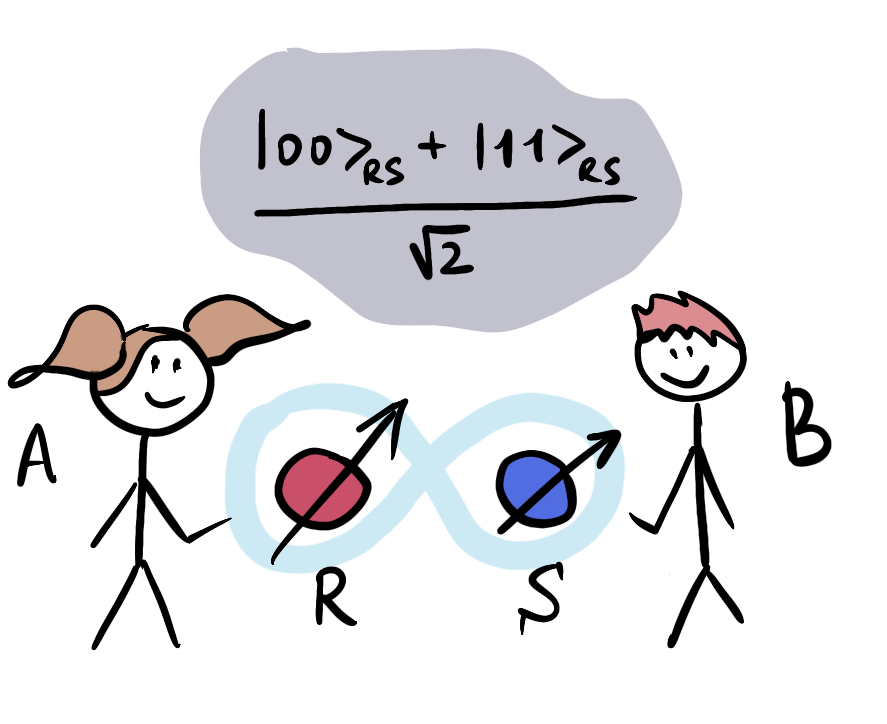}
    \caption{{\bf Alice and Bob measure a Bell state.} Alice and Bob  each have access to a half of a Bell state, and measure their own qubit in the computational basis. Alice should guess Bob's outcome.}
    \label{fig:ex1}
    \end{subfigure}
    \
    \begin{subfigure}{0.45\textwidth}
    \centering
    \includegraphics[scale=0.3]{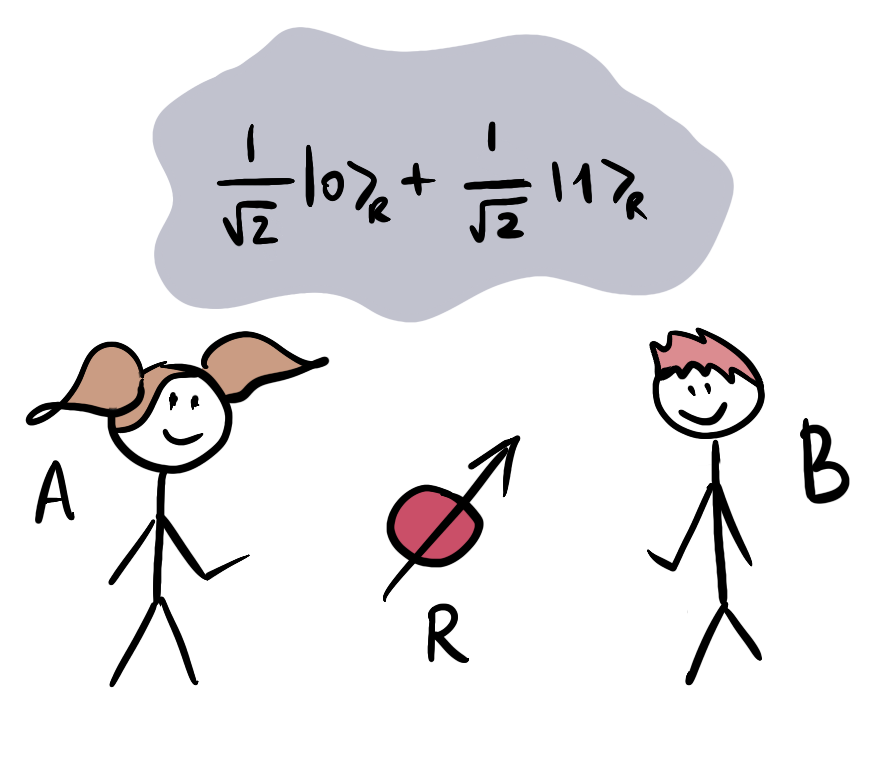}
    \caption{{\bf Alice and Bob measure the same qubit.} Alice and Bob share a qubit in state $\ket+$. Alice measures it, and then tries to predict the outcome of  Bob's subsequent measurement.}
    \label{fig:ex2}
    \end{subfigure}
    \\
    \begin{subfigure}{0.45\textwidth}
    \centering
    \includegraphics[scale=0.3]{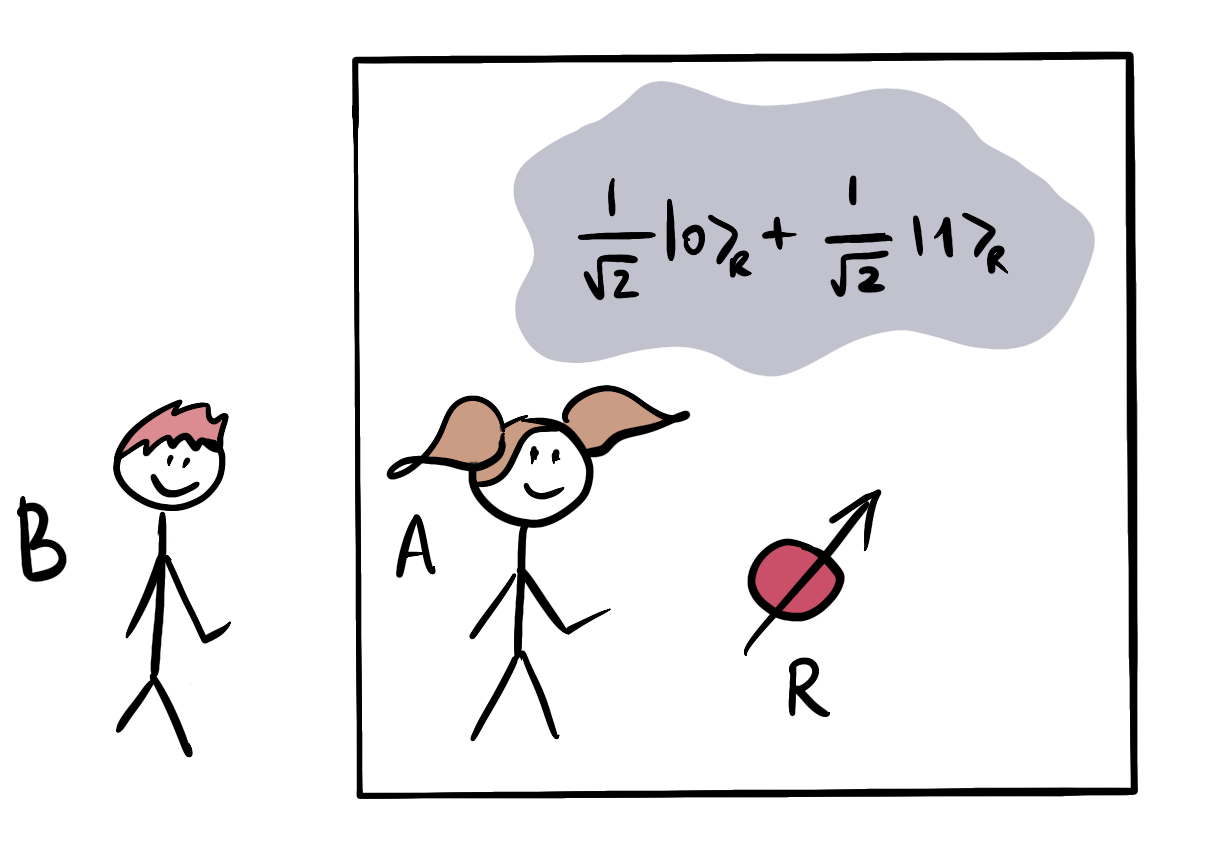}
    \caption{{\bf Bob measures Alice.} Alice measures a qubit (in state $\ket+$), and then tries to predict the outcome of  Bob's future measurement. Bob reverses Alice's memory update and measures her memory.}
    \label{fig:ex3}
    \end{subfigure}
    \caption{\textbf{Simplest test protocols.}  Here  two agents measure a shared state and reason about each other's outcomes; there are no measurements of each other's memories involved. These ``tutorial'' examples act as simple tests of whether an interpretation was implemented correctly in Quanundrum and makes the expected predictions in familiar settings.}
\end{figure}

\subsection{Alice and Bob measure a Bell state [\href{https://github.com/jangnur/Quanundrum/blob/master/quanundrum/simpleExamples/simple\%20example\%20I.ipynb}{Jupyter notebook}]}

\begin{figure}[t]
    \centering
    \begin{align*}
        \Qcircuit @C=1.3em @R=1.3em {
        \lstick{\textbf{R} \hspace{3.5em}} & \ctrl{2} & \qw \qw & \qw & \qw & \qw & \qw & \qw \\
        \lstick{\textbf{S} \hspace{3.5em}} & \qw & \qw & \qw & \qw & \qw & \ctrl{8} & \qw  \inputgroupv{1}{2}{0.7em}{0.9em}{\ket{\psi}_{SR} \hspace{1em}}  \\
        \lstick{\text{Alice's meas. }\ket 0_A} & \targ &  \meter \qw & \ustick{a} \cw & \multigate{6}{\text{Reasoning}} \cw & \cw & \cw & \cw \\
        \lstick{[a=0 \Rightarrow b=0] \quad \ket 0} & \qw & \qw & \qw & \ghost{\text{Reasoning}} & \qw & \qw & \qw \\
        \lstick{[a=0 \Rightarrow b=1] \quad \ket 0} & \qw & \qw & \qw & \ghost{\text{Reasoning}} & \qw & \qw & \qw \\
        \lstick{[a=1 \Rightarrow b=0] \quad \ket 0} & \qw & \qw & \qw & \ghost{\text{Reasoning}} & \qw & \qw & \qw \\
        \lstick{[a=1 \Rightarrow b=1] \quad \ket 0} & \qw & \qw & \qw & \ghost{\text{Reasoning}} & \qw & \qw & \qw \\
        \lstick{\text{Prediction } b=0 \quad \ket 0} & \qw & \qw & \qw & \ghost{\text{Reasoning}} & \qw & \qw & \qw \\
       \lstick{\text{Prediction } b=1 \quad \ket 0} & \qw & \qw & \qw & \ghost{\text{Reasoning}}& \qw & \qw & \qw \\
        \lstick{\text{Bob's meas. }\ket 0_B} & \qw & \qw & \qw & \qw & \qw & \targ & \meter \qw \\
        }
    \end{align*}
    \caption{{\bf Alice and Bob measure a Bell state: Alice's prediction-making.} Alice and Bob have access to qubits $R$ and $S$ respectively, which are initially prepared in Bell state $\ket{\psi}_{SR} = \frac{1}{\sqrt 2} (\ket{00}_{SR} + \ket{11}_{SR})$. Alice measures  qubit $R$, obtaining a classical outcome, which she writes down to her (classical) memory. For both of her outcome options ($a=0$ and $a=1$) she runs the circuit. Neither the inference nor the prediction qubits are  initialized at that point. After Bob's measurement of the system $S$, and looking up his measurement outcome, she is able to see if there is any logical correlation of his outcome with her own result.}
    \label{fig:bellAlice}
\end{figure}
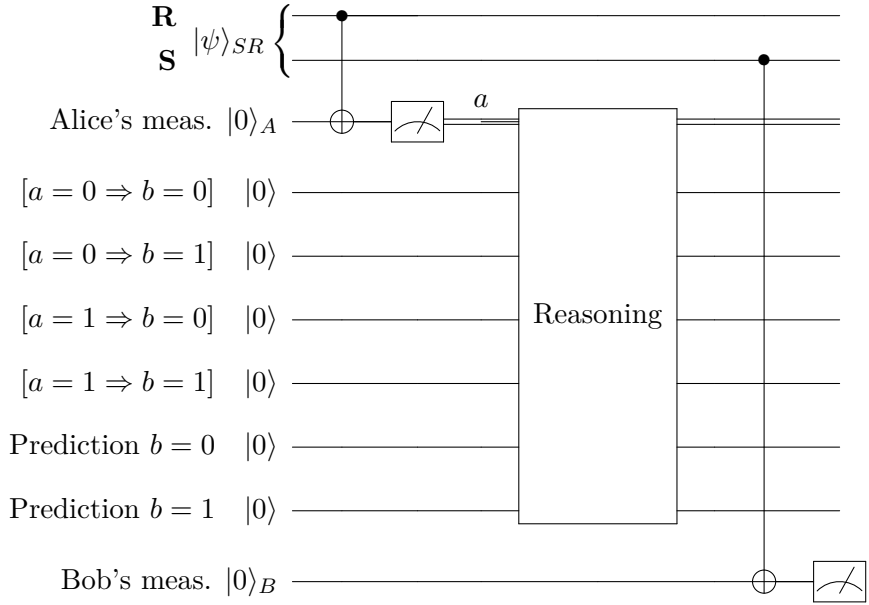

\paragraph{Setting.} 
Alice and Bob share a Bell pair $(\ket{00}_{RS}+\ket{11}_{RS})/\sqrt{2}$ (Figure~\ref{fig:ex1}). The experiment proceeds as follows:
\begin{itemize}
\item [$t=1$] Alice measures her qubit $R$ in basis $\{\ket{0}_R,\ket{1}_R\}$, and records the result in her memory $A$.
\item [$t=2$] Alice makes a prediction about Bob's outcome.
\item [$t=3$] Bob measures his qubit $S$ in basis $\{\ket{0}_S,\ket{1}_S\}$, and records the result in his memory $B$.
\end{itemize}

\paragraph{Expected result.} 
In most interpretations of quantum theory we expect Alice to correctly guess that their outcomes are perfectly correlated. 

\paragraph{Running the protocol.}
After defining the steps of agents as described above, we need to initialize Alice's predictions about Bob's state. From Alice's point of view, she runs the protocol with her own outcomes being classical, and the inference and prediction qubits not initialized. Her circuit of reasoning is pictured on Figure~\ref{fig:bellAlice}.

\paragraph{Forward reasoning.}
Using her knowledge of quantum theory, Alice can run a simulation of the whole experiment (before step $t_1$), and update her instruction registers to reflect the following: 
``if my measurement outcome is 0, I should predict that Bob will obtain 0; if my measurement outcome is 1, I should predict that Bob will obtain 1.''  
This can be economically encoded by initializing her prediction qubit to $\ket 0$ (the default prediction), setting her first instruction qubit to $\ket 0$ (``if I see 0, I should not change my prediction'') and her second instruction qubit to $\ket1$ (``if I see 1, transform the prediction''). When the experiment actually runs, one can simulate Alice's reasoning by running her circuit between steps $t_1$ and $t_2$, obtain her prediction, and correlate it  with Bob's outcome at step $t_3$. The protocol is run until Alice's measurement, and each outcome is analyzed separately by projecting the state into the subspace corresponding to Alice getting the said outcome. For example, here the analysis would be carried out in the following way.

\begin{itemize}
    \item Case $a=0$:
    \begin{enumerate}
        \item $RS$ is initialized in the Bell state $\frac{1}{\sqrt 2} \ket{00}_{RS} + \frac{1}{\sqrt 2} \ket{11}_{RS}$; Alice measures $R$ and writes the result down to her memory,
        \begin{gather*}
           \frac{1}{\sqrt 2} \ket{00}_{RS} \ket 0_A + \frac{1}{\sqrt 2} \ket{11}_{RS} \ket 1_A
        \end{gather*}
        
        \item The state is projected onto the subspace of Alice's memory corresponding to $a=0$, $\ket{00}_{RS} \ket 0_A$.~\footnote{Any prediction qubits associated with Alice's inference for $a=0$ will also be captured in this subspace; here we don't add them explicitly to make the explanation simpler. For the full state, see the inference making function output in the jupyter notebooks.}
        
        \item Bob measures the state  
        \begin{gather*}
            \ket{00}_{RS} \ket 0_A \ket 0_B.
        \end{gather*}
        
        \item Now Alice can conclude that if she gets $a=0$, Bob obtains $b=0$.
    \end{enumerate}
    
    \item Case $a=1$:
    \begin{enumerate}
        \item $RS$ is initialized in the Bell state $\frac{1}{\sqrt 2} \ket{00}_{RS} + \frac{1}{\sqrt 2} \ket{11}_{RS}$; Alice measures $R$ and writes the result down to her memory,
        \begin{gather*}
           \frac{1}{\sqrt 2} \ket{00}_{RS} \ket 0_A + \frac{1}{\sqrt 2} \ket{11}_{RS} \ket 1_A
        \end{gather*}
        
        \item The state is projected onto the subspace of Alice's memory corresponding to $a=1$, $\ket{11}_{RS} \ket 1_A$.
        
        \item Bob measures the state, and writes the result down to his memory,  
        \begin{gather*}
            \ket{11}_{RS} \ket 1_A \ket 1_B.
        \end{gather*}
        
        \item Now Alice can conclude that if she gets $a=1$, Bob obtains $b=1$.
    \end{enumerate}
\end{itemize}

\subsection{Alice and Bob make sequential measurements [\href{https://github.com/jangnur/Quanundrum/blob/master/quanundrum/simpleExamples/simple\%20example\%20II.ipynb}{Jupyter notebook}]}

This is an almost trivial experiment that is used as a quick test for new interpretations and logical axioms.

\paragraph{Setting.} 
Alice and Bob  have access to the same qubit $R$ , initially in state
 $\ket{+}_R$ (Figure~\ref{fig:ex2}). The experiment proceeds as follows:
\begin{itemize}
\item [$t=1$] Alice measures system $R$ in basis $\{\ket{0}_R,\ket{1}_R\}$, and records the result in her memory $A$.
\item [$t=2$] Bob measures system $R$ in basis $\{\ket{0}_R,\ket{1}_R\}$, and records the result in his memory $B$.
\item [$t=3$] Alice and Bob reason about each other's outcomes.
\end{itemize}

\paragraph{Expected result.} According to most standard quantum interpretations, we should expect their results to be the same, and them to be able to correctly guess each other's outcome; this is also our experience in the lab.

\subsection{Bob measures Alice [\href{https://github.com/jangnur/Quanundrum/blob/master/quanundrum/simpleExamples/simple\%20example\%20III\%20-\%20Copenhagen.ipynb}{Jupyter notebook}]}

\paragraph{Setting.}
Alice has access to a qubit $R$, and Bob has access to Alice's memory (Figure~\ref{fig:ex3}). The initial state of $R$ is $\frac{1}{\sqrt{2}} (\ket{0}_R + \ket{1}_R)$. The experiment proceeds as follows:
\begin{itemize}
\item [$t=1$] Alice measures system $R$ in basis $\{\ket{0}_R,\ket{1}_R\}$, and records the result in her memory qubit $A$.
\item [$t=2$] Alice makes a prediction about Bob's outcome at $t=4$.
\item [$t=3$] Bob applies a CNOT gate on Alice's lab, which is controlled on the state of the qubit $R$.
\item [$t=4$] Bob measures Alice's memory $A$.
\end{itemize}

\paragraph{Expected result.}
According to the neo-Copenhagen interpretation, Alice's prediction about Bob's outcome will be that he gets $0$, as Alice's memory qubit $A$ starts out in the state $\ket 0_A$. According to the collapse interpretation, Alice cannot make a deterministic prediction about Bob's outcome at a later step; however, Bob's measurement result corresponds to her state $\ket 0_A$.

%% file: FR_protocol.tex
\begin{figure}[t]
    \centering
    \includegraphics[scale=0.3]{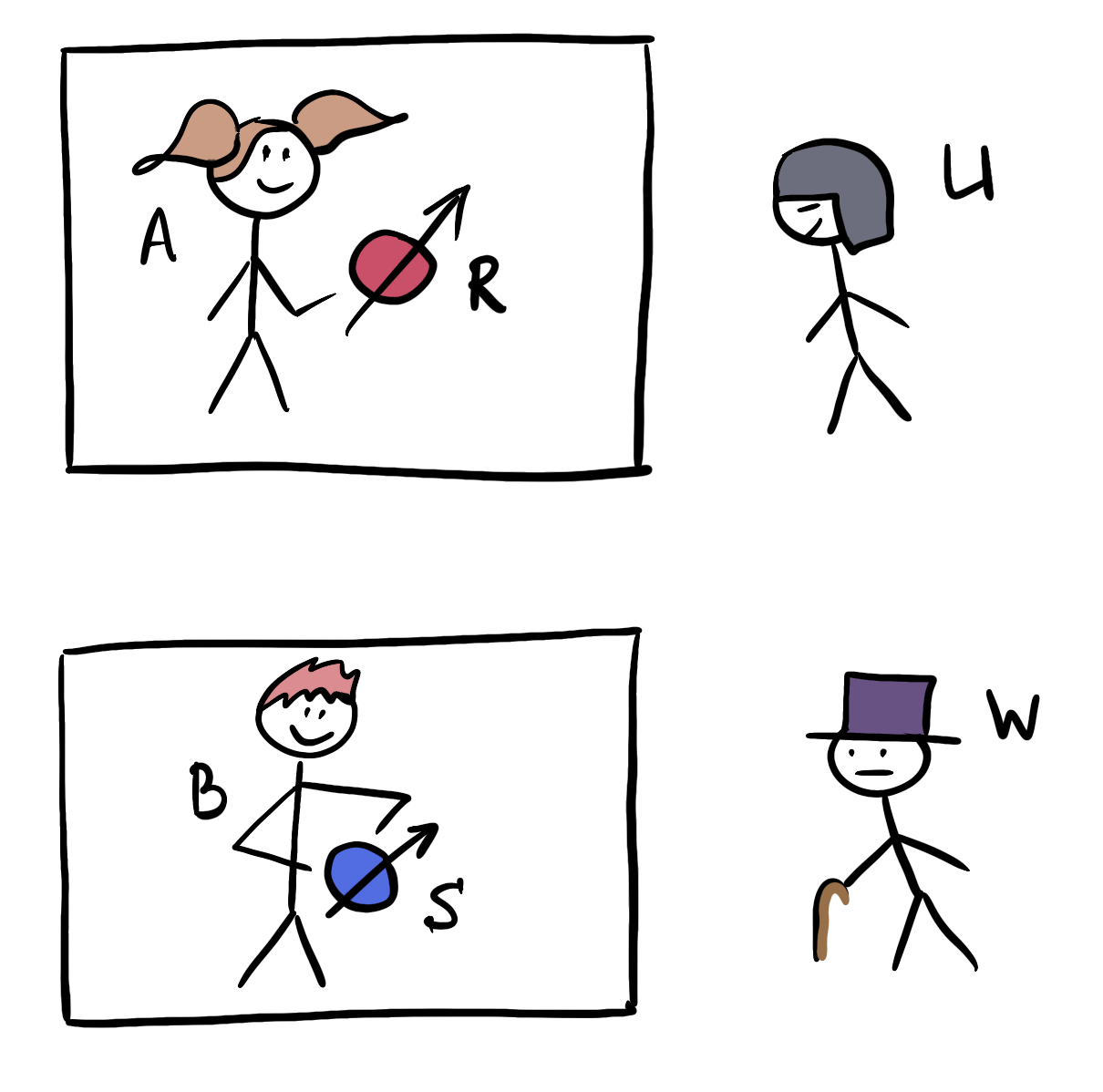}
    \caption{{\bf FR thought experiment~\cite{Frauchiger2018}.} The setting involves four agents: Alice, Bob, Ursula and Wigner, and two qubit systems $R$ and $S$. Alice measures system $R$ and prepares system $S$ based on her outcome. The system $S$ is then measured by Bob. Ursula and Wigner measure Alice's and Bob's labs respectively. The agents are allowed to make predictions with certainty about each other's outcomes.}
    \label{fig:fr}
\end{figure}

We will now see an application of all the machinery described in the previous sections, by simulating the Frauchiger-Renner thought experiment. \href{https://github.com/jangnur/Quanundrum/tree/master/quanundrum/FrauchigerRennerExample}{Jupyter notebooks} are available for the implementation with the
 neo-Copenhagen interpretation  and  a collapse interpretation, and we recommend following them.
We will first go through the interpretation with the neo-Copenhagen interpretation, and later comment on the collapse version. 

\paragraph{Global setting.} 
The full thought experiment involves four agents, Alice, Bob, Ursula and Wigner, and two qubits $R$ and $S$ (Figure~\ref{fig:fr}).
We start from the example we have followed in the main text (with Alice, Bob, $R$ and $S$), and introduce one new agent at the time. For full generality, we assume that every agent may be modelled as a quantum reasoning agent of $n_i$ qubits  (an outcome register $i$, inference qubits $I_i$ and prediction registers $P_i$), and a unitary reasoning circuit $U_i$. We assume that the reasoning process of all agents is common knowledge (all agents know $\{U_i\}_i$ for all other agents).

\subsection{Introducing Ursula: a meta agent who measures Alice}

\begin{figure}[t]
    \Qcircuit @C=1.3em @R=1.3em {         
       \push{\hspace{11em}} &  &  & \blue{\mbox{Alice's actions}}  & & & & & &   \blue{\mbox{Ursula's actions}}   \\
       &\lstick{\textbf{Qubit $R$: } \sqrt{\frac13}\ \ket 0_R + \sqrt{\frac23}\  \ket1_R}     & \ctrl{1} &  \qw &  \qw  &  \qw & \qw & \qw & \qw & \qw &  \multimeasureD{1}{\text{Bell}} &  \dstick{\green{=u} \quad }   \\
         &\lstick{\textbf{Alice's outcome: } \ket 0_A} 
            & \targ & \ctrl{2} &  \multigate{1}{U_\text{Alice}}   \qw &  \qw  &  \qw & \qw & \qw &  \multigate{1}{U_\text{Alice}^{-1}} & \ghost{\text{Bell}}  &  \\
        &\lstick{\textbf{Alice's brain: } \ket{\vec x_A}_{I_A} \ket{0}_{P_A}} 
            & \qw &  \qw & \ghost{U_{\text{Alice}}}  &  \qw  &   \qw & \dstick{\blue{\mbox{Bob's actions}}}  \qw & \qw &  \ghost{U_\text{Alice}^{-1}} & \qw &  \\
        &\lstick{\textbf{Qubit $S$: } \ket 0_S}
           &  \qw & \gate{H} & \qw  &  \qw &   \ctrl{1}  & \qw  & \qw  \\
        &\lstick{\textbf{Bob's outcome: } \ket 0_B} 
            &  \qw &  \qw  &  \qw & \qw &  \targ & \multigate{3}{U_\text{Bob}} & \qw \\
        &\lstick{\textbf{Bob's inference: }  \ket {\vec{x_B}}_{I_B}}
           &  \qw &  \qw  &  \qw & \qw & \qw & \ghost{U_\text{Bob}}  & \qw  \\
        &\lstick{\textbf{Bob's prediction: } a=0 \quad \ket 0_{P_B^0} }
           &  \qw &  \qw  &  \qw & \qw & \qw &  \ghost{U_\text{Bob}}& \qw  \\
        &\lstick{\textbf{Bob's prediction: } a=1 \quad \ket 0_{P_B^1}}
            &  \qw &  \qw  &  \qw & \qw & \qw &  \ghost{U_\text{Bob}}& \qw 
         \gategroup{2}{3}{5}{6}{.7em}{--} 
        \gategroup{5}{7}{9}{9}{.7em}{--} 
        \gategroup{2}{10}{4}{12}{.7em}{--}
    }
    \caption{{ \bf Global circuit, from Ursula's perspective, up to Ursula's measurement.}  Alice and Bob's brains are simulated by $n_A$ and $n_B$ qubits respectively, including inference and prediction registers. Ursula's task is to  guess the content of Bob's two prediction register, given the outcome $u$ of her own measurement. Note that Alice's potential reasoning and predictions implemented through  Alice's reasoning circuit $U_{\text{Alice}}$, are reversed by Ursula. Bob's reasoning circuit $U_{\text{Bob}}$ is the one from Figure~\ref{fig:bob-circuit-full}. Ursula's measurement is in the Bell basis.}
    \label{fig:alice-bob-ursula-circuit}
\end{figure}
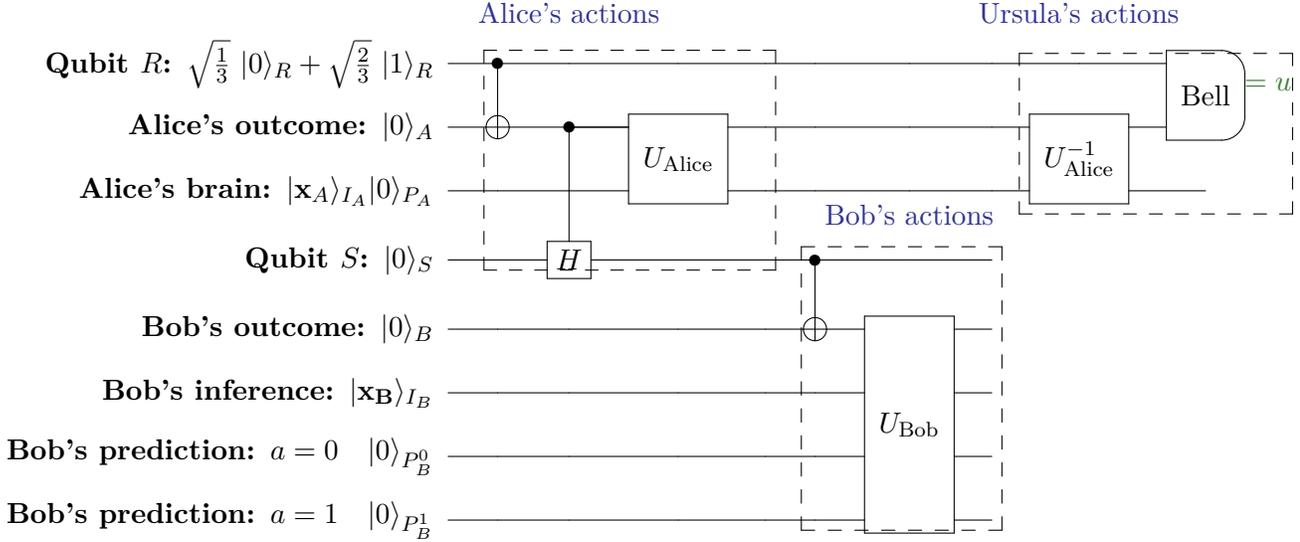

Ursula is an agent who stands outside of Alice's lab and has (hypothetically) full quantum control over all systems in that lab, including Alice's memory. 

\paragraph{Experimental setting.} To the previous experiment, we add two more steps:
\begin{itemize}
    \item[$t=2.9$] Ursula reverses Alice's unitary reasoning circuit (if present).
    \item[$t = 3$] Ursula
    measures part of Alice's lab:  qubit $R$ and Alice's output qubit register $A$. She measures them in the Bell basis. For future convenience, we name two states of this basis\footnote{The other two basis states can be given any name --- they don't have overlap with the state of the experiment and have probability zero. } as
\begin{gather*}
\ket{\text{ok}}_{RA}=\sqrt{\frac{1}{2}}(\ket{0}_R\ket{0}_A-\ket{1}_R\ket{1}_A)\\
\ket{\text{fail}}_{RA}=\sqrt{\frac{1}{2}}(\ket{0}_R\ket{0}_A+\ket{1}_R\ket{1}_A).
\end{gather*} 
    
    \item[$t=3.5$]  Ursula tries to guess what Bob's guess was at time $t=2.5$. 
\end{itemize}

\begin{figure}[t!]
\begin{subfigure}{\textwidth}
    \Qcircuit @C=1.3em @R=1em {     
        \push{\hspace{11em}} 
        &\lstick{\textbf{Qubit $R$: } \sqrt{\frac13}\ \ket 0_R + \sqrt{\frac23}\  \ket1_R}     & \ctrl{1} &  \qw & \qw &  \qw & \qw & \qw & \qw &  \multimeasureD{1}{\text{Bell}} &    \\
        &\lstick{\textbf{Alice's outcome: } \ket 0_A} 
            & \targ & \ctrl{2} &  \multigate{1}{U_\text{Alice}}   \qw &  \qw & \qw& \qw &   \multigate{1}{U_\text{Alice}^{-1}} & \ghost{\text{Bell}}  & \ustick{\blue{=u}} \\
        &\lstick{\textbf{Alice's brain: } \ket{\vec x_A}_{I_A}\ket0_{ P_A}}
            & \qw &  \qw & \ghost{U_{\text{Alice}}}   &  \qw & \ustick{ \green{\mathbbm 1_{\text{Alice}}}} \qw& \qw &  \ghost{U_\text{Alice}^{-1}} &  \qw \\
        &\lstick{\textbf{Qubit $S$: } \ket 0_S}
           &  \qw & \gate{H} & \qw  &   \ctrl{1}& \qw & \qw    \\
        &\lstick{\textbf{Bob's outcome: } \ket 0_B} 
            &  \qw &\qw  &  \qw  \qw &  \targ & \multigate{3}{U_\text{Bob}} & \qw \\
        &\lstick{\textbf{Bob's inference: }  \ket {0001}_{I_B}}
           &  \qw  &  \qw & \qw & \qw & \ghost{U_\text{Bob}} &\qw  \\
        &\lstick{\textbf{Bob's pred. } a=0 \quad \ket 0_{P_B^0} }
           &  \qw  &  \qw & \qw & \qw &  \ghost{U_\text{Bob}}&   \meter & \rstick{\blue{p_B^0}} \cw  \\
        &\lstick{\textbf{Bob's pred. } a=1 \quad \ket 0_{P_B^1}}
           &  \qw  &  \qw & \qw & \qw &  \ghost{U_\text{Bob}}&  \meter & \rstick{\blue{p_B^1}}  \cw
        \gategroup{2}{5}{3}{9}{.7em}{--} 
        }
    \caption{{\bf Simulating Bob's quantum brain.} Ursula reasons about Bob's prediction (of Alice's outcome) by  including Bob's brain in her simulation, and measuring Bob's prediction registers.  From the correlations between outcomes $u, p_B^0$ and $p_B^1$, she finds that the inference $u= \text{ok} \implies p_B^1=1$ is valid: if Ursula obtains outcome ``ok'', she knows that Bob will have predicted $a=1$.}
    \label{fig:alice-bob-ursula-simulation-large}
    \end{subfigure}
    \begin{subfigure}{\textwidth}
    \vspace{6mm}
     \Qcircuit @C=1.3em @R=1em {     
        \push{\hspace{15em}} 
        &\lstick{\textbf{Qubit $R$: } \sqrt{\frac13}\ \ket 0_R + \sqrt{\frac23}\  \ket1_R}     & \ctrl{1} &  \qw & \qw &  \qw & \qw & \qw &  \multimeasureD{1}{\text{Bell}} &    \\
         &\lstick{\textbf{Alice's outcome: } \ket 0_A} 
            & \targ & \ctrl{2} &  \multigate{1}{U_\text{Alice}}   \qw &  \qw & \qw&   \multigate{1}{U_\text{Alice}^{-1}} & \ghost{\text{Bell}}  & \ustick{\blue{= u}} \\
        &\lstick{\textbf{Alice's brain: } \ket{\vec x_A}_{I_A}\ket0_{ P_A}} 
            & \qw &  \qw & \ghost{U_{\text{Alice}}}   &  \qw & \ustick{ \green{\mathbbm 1_{\text{Alice}}}} \qw& \ghost{U_\text{Alice}^{-1}} &  \qw \\
        &\lstick{\textbf{Qubit $S$: } \ket 0_S}
           &  \qw & \gate{H} & \qw  &   \ctrl{1}& \qw   \\
        &\lstick{\textbf{Bob's outcome: } \ket 0_B} 
            &  \qw &\qw  &  \qw  \qw &  \targ &   \meter & \rstick{\blue{b}} \cw  
        \gategroup{2}{5}{3}{8}{.7em}{--} 
    }
    \caption{{ \bf Shorter simulation, classical logic.} Ursula only simulates Bob's lab up to his measurement, and then measures his simulated outcome registry directly to obtain the value of $b$. Through theoretical or statistical analysis, Ursula concludes that $u=$ok $\implies b=1$. Combining this classical inference with the previous inference $b=1 \implies a=1$, Ursula obtains $u=$ok $\implies a=1$.  }
    \label{fig:alice-bob-ursula-simulation-small}
    
    \end{subfigure}
    \caption{{\bf Ursula's restricted simulation.}  Ursula has two options to calculate the truth value of inferences $\{ u \Rightarrow a\}_{a,b}$ about Alice's past outcome based on her own observations. Note that  Ursula can safely ignore Alice's reasoning: her brain's circuit $U_{\text{Alice}}$ is reversed by Ursula in the actual experiment (so Ursula does not need to simulate it), and the initial state $\ket{\vec x_A}_{I_A}$ of Alice's inference registers is irrelevant for either simulation's statistics.}
    \label{fig:ursula_simulations}
\end{figure}
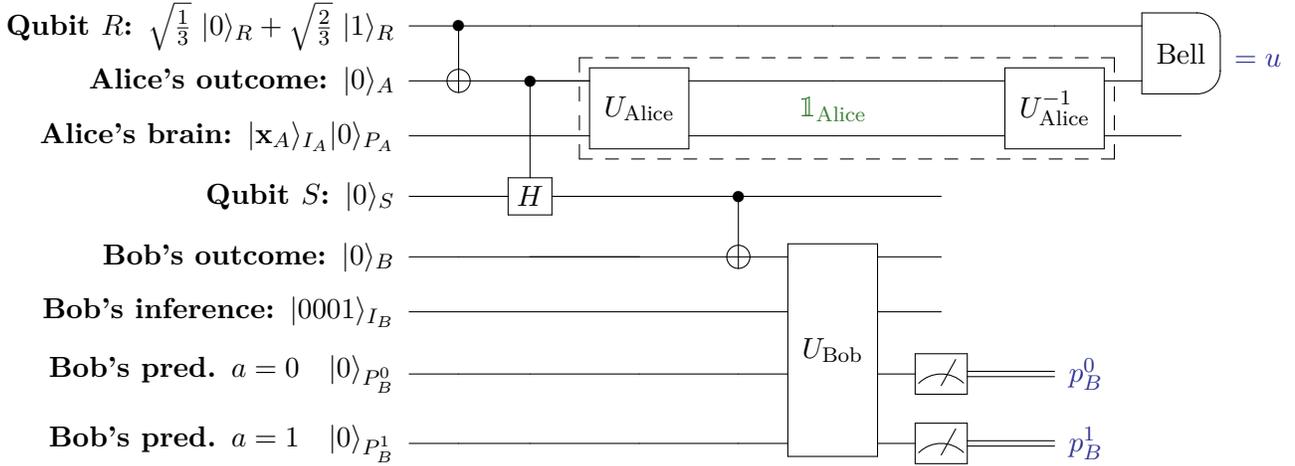
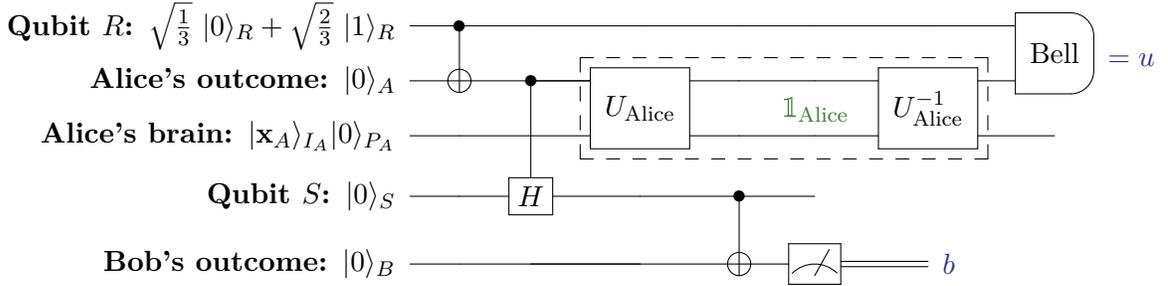

\paragraph{Ursula's first simulation.} From Ursula's perspective, the circuit representation of the experiment up to her measurement is given in  Figure~\ref{fig:alice-bob-ursula-circuit}. In addition, Ursula has her own reasoning circuit, similar to Bob's. In order to make accurate predictions, she will have to run several simulations. As a first step, she needs to initialize Bob's inference qubits $\ket{\vec x_B}_{I_B}$: for this she puts herself in Bob's place, and runs the restricted simulation described in the previous section (Figure~\ref{fig:bob-restricted-simulation}, with just Alice and Bob). After this, she knows that $\ket{\vec x_B}_{I_B} = \ket{0001}_{I_B}$, like we saw before. Note that even if we have an explicit circuit for Alice's reasoning, it should not affect the outcomes of this simulation, because it is reversed by $U_{\text{Alice}}^{-1}$. Next Ursula must relate the outcomes of her own measurement to Bob's predictions about Alice's outcomes. We have two options to achieve this, which can be easily implemented (Figure~\ref{fig:ursula_simulations}). 

\paragraph{First option: large, direct quantum simulation.} Ursula can run a simulation of the whole experiment up to now, with Bob's inference qubits initialized, and then measure the simulated Bob's  prediction registers directly (Figure~\ref{fig:alice-bob-ursula-simulation-large}). She compares the outcome $u$ of her Bell measurement on Alice's lab to the outcomes $p_B^0$ and $p_B^1$ of Bob's prediction registers. In this case, she'll obtain the result that whenever $u=$ok, then $p_B^0=0, p_B^1=1$; that is, Ursula concludes that when she measures outcome ``ok'', Bob will have predicted that $a=1$. 

\paragraph{Second option: shorter simulation of Bob's memory.} Alternatively Ursula can run a shorter simulation to infer directly Bob's  measurement outcome $b$ (Figure~\ref{fig:alice-bob-ursula-simulation-small}). This is essentially a four-qubit simulation, without modelling the inner circuits of Alice's or Bob's brains, so this method is computationally much more  economical for Ursula. From it, she learns the inference $u=$ok $\implies b=1$. The reader can see this directly from  the joint state of qubits  $RASB$ after Alice's reasoning is undone,
\begin{align*}
    \ket{\psi(t=3)}_{RASB} 
    &=\frac1{\sqrt3} \left(  \ket{00}_{RA}\ket{00}_{SB} + \ket{11}_{RA}\ket{00}_{SB}  +\ket{11}_{RA}\ket{11}_{SB}  \right) \\
    &= \sqrt{\frac23} \ket{\text{fail}}_{RA} \ket{00}_{SB} 
    + \sqrt{\frac16} \left(\ket{\text{fail}}_{RA}  + \ket{\text{ok}}_{RA} \right) \ket{11}_{SB} . 
\end{align*}
Here the only outcome in $B$ that has non-zero overlap with $u=$ok is $b=1$. 
The price to pay for this cheaper simulation is heavier reliance on classical logic axioms: Ursula must classically  combine $u=$ok $\implies b=1$ with the previous inference $b=1\implies a=1$ by using the knowledge distribution axiom, to obtain $u=$ok $\implies a=1$.\footnote{Note that this diverges slightly  from Ursula's deduction in the original FR paper, where she waits until the end to compute a prediction for Wigner's outcome $w$ directly.}  

\paragraph{The fine print.} In the software, the user can choose which version to run by specifying in the protocol the target registers of Ursula's reasoning at time $t=3.5$.  In the first case, Ursula reasons about qubits $P_B^0$ and $P_B^1$, and in the shorter simulation she reasons about $B$. Formally, in the first simulation, the epistemic statement that Ursula reaches is $K_U ( [u=\text{ok}]  \implies K_B [a=1] )$. Through the second option, she reaches  $K_U ( [u=\text{ok}]  \implies [b=1] )$  and $K_U (K_B [b=1 \implies a=1])$. Whether she is allowed to combine those statements depends on the logical axioms and trust structures applied by the user.

\subsection{Introducing Wigner: a meta agent who measures Bob's lab}

Analogously to Ursula, Wigner is an agent who stands outside of Bob's lab and has full quantum control over all systems in that lab, including Bob's memory. 

\begin{figure}[t]
 \Qcircuit @C=1.3em @R=1.3em { 
    \push{\hspace{6em}}  & & &  \mbox{ \blue{Alice's actions}} & &  & & & &  \mbox{\blue{Ursula's actions}}      \gategroup{2}{3}{5}{6}{.7em}{--} 
        \gategroup{5}{7}{7}{8}{.7em}{--} 
        \gategroup{2}{9}{4}{11}{.7em}{--}
        \gategroup{5}{11}{7}{14}{.7em}{--}
        \\
        &\lstick{\textbf{$R$: } \sqrt{\frac13}\ \ket 0_R + \sqrt{\frac23}\  \ket1_R}     & \ctrl{1} &  \qw &  \qw &  \qw  & \qw & \qw  & \qw &  \multimeasureD{1}{\text{Bell}} & \dstick{\green{=u} \qquad}   & &   \\
        &\lstick{\textbf{A's outcome: } \ket 0_A} 
            & \targ & \ctrl{2} &  \multigate{1}{U_\text{Alice}}  &  \qw &   \qw & \qw &  \multigate{1}{U_\text{Alice}^{-1}} & \ghost{\text{Bell}} &    \\
         &\lstick{\textbf{A's brain: } \ket{\vec x_A}_{I_A} \ket{0}_{P_A}} 
            &   \qw &  \qw & \ghost{U_{\text{Alice}}} &  \qw   &  \qw & \dstick{\mbox{\blue{Bob's actions} \quad }}    \qw &  \ghost{U_\text{Alice}^{-1}} & \qw & & \dstick{\mbox{ \blue{Wigner's actions}}} \\
       &\lstick{\textbf{$S$: } \ket 0_S}
           &  \qw & \gate{H} & \qw &  \qw  &   \ctrl{1}  & \qw & \qw & \qw &  \qw  &\multimeasureD{1}{\text{Bell}}  & & \\
        &\lstick{\textbf{B's outcome: } \ket 0_B} 
            &  \qw &  \qw &\qw  &  \qw  \qw &  \targ & \multigate{1}{U_\text{Bob}} & \qw &  \qw & \multigate{1}{U^{-1}_\text{Bob}} &  \ghost{\text{Bell}} & \ustick{\green{= w}}  &  \\
        &\lstick{\textbf{B's brain: }  \ket {\vec{x_B}}_{I_B} \ket 0_{P_B} }
           &  \qw &  \qw  &  \qw & \qw & \qw & \ghost{U_\text{Bob}}  &  \qw& \qw &\ghost{U^{-1}_\text{Bob}}   & \qw &  &
  }
  \caption{{ \bf Circuit of the full experiment, from Ursula and Wigner's perspective.} After Ursula's measurement, Wigner reverses Bob's reasoning and measures his lab, obtaining outcome $w$. If $u=w = \text{ok}$ (which happens with probability $1/12$), they reason about Alice's past prediction of Wigner's outcome.  }
    \label{fig:alice-bob-ursula-wigner-circuit}
\end{figure}
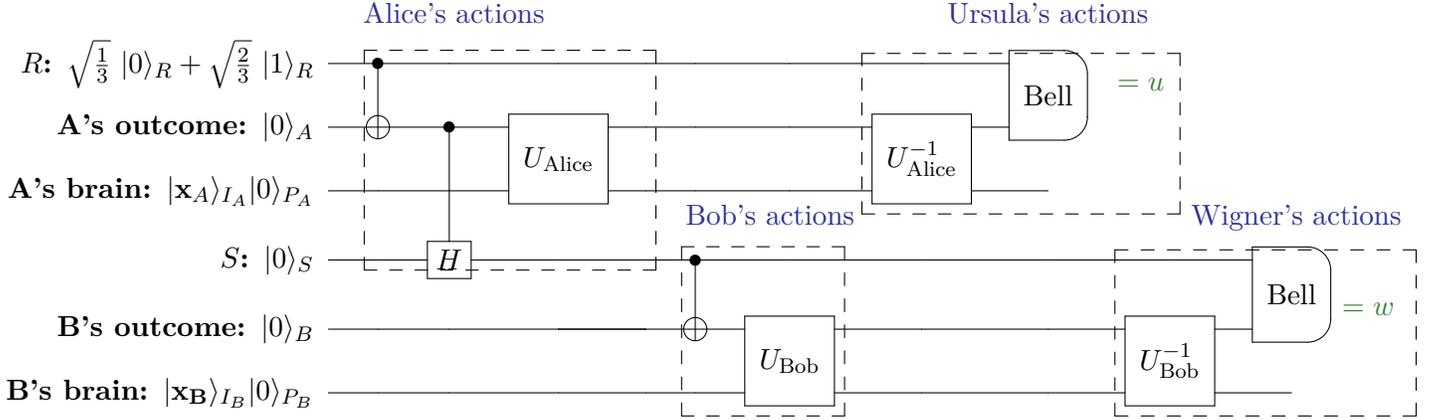

\paragraph{Experimental setting.} We add a few more steps to the protocol (Figure~\ref{fig:alice-bob-ursula-wigner-circuit}). Note that one of the new  steps specifies Alice's reasoning early in the experiment:
\begin{itemize}
    \item[$t=3.9$] Wigner reverses Bob's unitary reasoning circuit (if present).
    \item [$t=4$] Wigner
    measures part of Bob's lab:  qubit $S$ and Bob's output qubit register $B$. He measures them in the Bell basis. Let us again name two states of this basis as
\begin{gather*}
\ket{\text{ok}}_{SB}=\sqrt{\frac{1}{2}}(\ket{0}_S\ket{0}_B-\ket{1}_S\ket{1}_B)\\
\ket{\text{fail}}_{SB}=\sqrt{\frac{1}{2}}(\ket{0}_S\ket{0}_B+\ket{1}_S\ket{1}_B).
\end{gather*} 
    \item [$t=1.6$] Alice tries to guess the outcome $w$ of Wigner's measurement at $t=4$.\footnote{The exact time stamp does not matter, only the relative order of operations.}

    \item [$t=5$] Ursula and Wigner compare the outcomes of their measurements. If they were both ``ok'', they halt the experiment and try to guess Alice's prediction at $t=1.5$.   Otherwise, they reset the timer and all systems to the initial conditions, and repeat the experiment. 
\end{itemize}

\begin{figure}[t]
    \Qcircuit @C=1.3em @R=1.3em {   
      \push{\hspace{11em}} 
      & \lstick{\sqrt{\frac13}\ \ket 0_R + \sqrt{\frac23}\  \ket1_R}   &  \ctrl{1} &  \qw & \qw &  \qw & \qw &    \\
         &\lstick{\textbf{Alice's outcome: } \ket 0_A} 
            & \targ  & \meter&   \ustick{\blue{ \ket{\psi_a}}}  \qw &   \ctrl{1}  &  \qw   \\
         &\lstick{\textbf{Qubit $S$: } \ket 0_S}
            & \qw&  \ustick{\mbox{\blue{project on $a$}}} \qw  &\qw    & \gate{H}  & \ctrl{1} & \qw & \qw & \qw& \multimeasureD{1}{\text{Bell}} &\dstick{\blue{=w}}  \\
        &\lstick{\textbf{Bob's outcome: } \ket 0_B} 
            &  \qw &\qw  &  \qw & \qw &  \targ & \multigate{1}{U_\text{Bob}} & \qw  & \multigate{1}{U^{-1}_\text{Bob}} &  \ghost{\text{Bell}}  &  \\
        &\lstick{\textbf{Bob's brain: }  \ket {\vec{x_B}}_{I_B} \ket 0_{P_B} }
           &  \qw  & \qw & \qw & \qw & \qw & \ghost{U_\text{Bob}} & \ustick{ \green{\mathbbm 1_{\text{Bob}}}}  \qw &\ghost{U^{-1}_\text{Bob}} & \qw
    \gategroup{4}{8}{5}{10}{.7em}{--} 
  }
    \caption{{ \bf Alice's restricted simulation.} In order to determine the validity of the various possible inferences $a\implies w$, Alice runs a restricted simulation of the experiment. She simulates a measurement on her memory and proceeds with independent simulations with the post-measurement states corresponding to each of her outcomes. Alice could alternatively run a larger simulation including  the reasoning circuits in her own brain and Ursula's undoing of those; this would lead to the same conclusion, $a=1 \implies w=$fail.    }
    \label{fig:alice-restricted-simulation}

\end{figure}
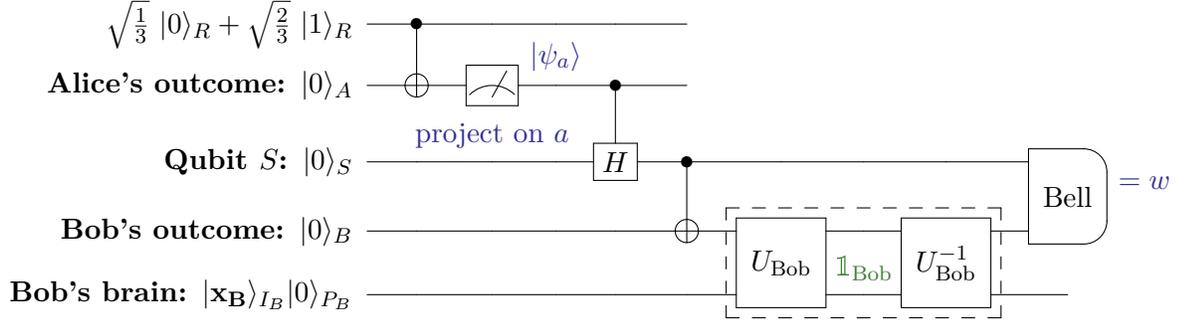

\paragraph{Alice's simulation.} Alice will have to make a prediction about whether Wigner obtains $w=\text{ok}$ or  $w=\text{fail}$, based on her observed outcome $a \in \{0,1\}$. This means that other agents will model her brain including two prediction qubits $P^{\text{ok}}$ and $P^{\text{fail}}$, and four inference qubits $\{I^{a,w}\}_{a,w}$, initially storing  the truth values for the different possibilities $\{a \Rightarrow w\}_{a,w}$. Her reasoning circuit is identical to Bob's (one outcome qubit, four inference and two predictions, with the four doubly-controlled gates).
To initialize her inference qubits, Alice runs a restricted version of the experiment, and similarly to the previous case, she has a few options for what to include in her simulation (for example whether to include her own brain and Ursula's measurement). Since in the simulation Alice projects on the post-measurement state for each of her outcomes $a$, these options lead to the same results; we present  the simplest version in Figure~\ref{fig:alice-restricted-simulation}. After running this simulation, Alice concludes that $a=1 \Rightarrow w = \text{fail}$, because for $a=1$ she projects $RA$  to $\ket1_R \ket1_A$, which leads to the global evolution 
\begin{align*}
    &\ket1_R\  \ket1_A \  \ket0_S \ \ket0_B  \ \ket{\vec x_B}_{I_B} \ket0_{P_B} \\
   \blue{\operatorname{CH}_{AS}} \longrightarrow & 
    \ket1_R \ \ket1_A\  \ket+_S \ \ket0_B \ \ket{\vec x_B}_{I_B} \ \ket0_{P_B}\\ 
    \blue{\operatorname{CNOT}_{SB}} \longrightarrow & 
    \ket1_R\  \ket1_A \   \frac{\ket0_S \ket0_B + \ket1_S \ket1_B}{\sqrt2} \ \ket{\vec x_B}_{I_B} \ \ket0_{P_B}\\ 
    \blue{U^{-1}_{\text{Bob}}\ U_{\text{Bob}}  }  \longrightarrow  & 
    \ket1_R \ \ket1_A  \ \green{\underbrace{\frac{\ket0_S \ket0_B + \ket1_S \ket1_B}{\sqrt2} }_{\ket{\text{fail}}_{SB}}} \ \ket{\vec x_B}_{I_B}\  \ket0_{P_B}.
\end{align*}
All the other possible inferences are false. Therefore we conclude that her inference qubits start in state $\ket{\vec x_A}_{I_A} = \ket{0001}_{I_A}$, with the last qubit corresponding to $\ket1_{I^{1,\text{fail}}}$. Note that other agents can put themselves in Alice's shoes and run the same simulation.

\paragraph{Wigner and Ursula's simulations.} Now Wigner and Ursula have several options for the simulations they run in order to conclude that if they observe $u=w=\text{ok} $ then Alice will have predicted $w=\text{fail}$ with certainty. The computationally lighter option is to classically combine all the inferences $u=\text{ok} \Rightarrow b=1$, $b=1 \Rightarrow a=1 $ and $a=1 \Rightarrow u = \text{fail}$; they can do this if the user-specified logical axioms allow for knowledge distribution among agents. An alternative that is computationally heavier but does not rely so much on classical logic is to run larger simulations. The software runs the following procedure automatically (we just need to specify which qubits agents are reasoning about). The circuit diagrams for every individual the simulation are quite large and repetitive, so we are not drawing them, but here is a compressed description:
\begin{itemize}
    \item[1.] They simulate Alice's reasoning (as per Figure~\ref{fig:alice-restricted-simulation}), to find the inference $a=1 \Rightarrow w= \text{fail}$ initial states for her inference registers, $\ket{\vec x_A}_{I_A} = \ket{0001}_{I_A}$. 
    
    \item[2.] Then they simulate Bob's reasoning about Alice's reasoning. That is, we think that Bob wants to guess not Alice's outcome $a$ but the content of her two prediction registers $P_A$. For economical reasons, we can just focus on Bob's thoughts about the one prediction qubit that is relevant for the logical contradiction, $P_A^{\text{fail}}$.\footnote{To reason about Alice's two prediction qubits, we'd need to double Bob's brain size (four prediction qubits $\{P_\text{ok}^0, P_\text{ok}^1, P_\text{fail}^0,P_\text{fail}^1\} $ and eight possible inferences), which the software does automatically, but might get out of hand for practical simulations. Alternatively we can find a compressed version by changing Bob's inner circuitry.}  
    In this case, Bob's prediction qubits will be $P^{\text{Bob  cannot guarantee that Alice predicted fail}}$ and $P^{\text{Bob knows that  Alice predicted fail with certainty}}$. 
    To follow Bob's reasoning, they must simulate Bob simulating Alice, so that they can initialize Bob's inference registers.  
    This is done through an expanded version of the circuit of Figure~\ref{fig:bob-restricted-simulation}, now including Alice's brain explicitly. Instead of measuring Alice's outcome registers, simulated Bob measures her prediction registers (analogously to Figure~\ref{fig:alice-bob-ursula-simulation-large}).  One deterministic conclusion of this simulation is $b=1 \Rightarrow p_A^{\text{fail} }=1$, that is: when Bob observes $b=1$ he knows that Alice will have concluded that Wigner would obtain outcome $w=\text{fail}$.  Now Ursula and Wigner can initialize Bob's inference qubits, in particular the qubit that encodes this inference, $\ket{1}_I^{b=1 \Rightarrow \text{Bob knows that  Alice predicted fail with certainty}}$ . 
    
    \item[3.] Next they need to find out Bob's prediction given Ursula's outcome $u$: Ursula runs a simulation similar to Figure~\ref{fig:alice-bob-ursula-simulation-large}; the difference is that now she measures Bob's prediction registers that store his conclusions about Alice's prediction (not her outcome $a$). The conclusion is that $u=1 \Rightarrow \text{Bob knows that  Alice predicted fail with certainty} $.  Now Ursula can initialize her own inference qubits with what she learned about Bob's knowledge of Alice's prediction, including the critical $\ket{1}_I^{u=1 \Rightarrow \text{Ursula knows that Bob knows that  Alice predicted fail with certainty}}$. 
    \item[4.] Finally we can run the whole experiment, with Alice, Bob and Ursula's brains initialized. We don't need to model Wigner's brain explicitly, just his outcome register $W$. At the end we measure the registers storing $w, u$ and Ursula's predictions. We find that whenever $u=w= \text{ok}$, Ursula's memory stores the chain of reasoning leading to Alice's prediction that $w= \text{fail}$.
\end{itemize}



